\documentclass[
prd,
twocolumn,
showpacs,
amsfonts,
amsmath,
amssymb,
nobibnotes,
nofootinbib,
eqsecnum,
]{revtex4}
\usepackage{type1cm,bm,dcolumn}
\usepackage[dvips]{graphicx,color}
\begin{document}


\title{Localizing gravity on Maxwell gauged $\mathbb{C}P^{1}$ model in six dimensions}
\author{Yuta Kodama}
\email{j6207613(at)ed.noda.tus.ac.jp}
\author{Kento Kokubu}
\author{Nobuyuki Sawado}
\email{sawado(at)ph.noda.tus.ac.jp}
\affiliation{Department of Physics, Faculty of Science and Technology, 
Tokyo University of Science, Noda, Chiba 278-8510, Japan}
\date{\today}


\begin{abstract}
We shall consider about a 3-brane embedded in six-dimensional space-time with a negative bulk cosmological constant.
The 3-brane is constructed by a topological soliton solution living in two-dimensional axially symmetric transverse subspace.
Similar to most previous works of six-dimensional soliton models,
our Maxwell gauged $\mathbb{C}P^{1}$ brane model can also achieve to localize gravity around the 3-brane.
The $\mathbb{C}P^{1}$ field is described by a scalar doublet and derived from $O(3)$ sigma model by projecting it onto two-dimensional complex space.
In that sense,
our framework is more effective than other solitonic brane models concerning with gauge theory.   
We shall also discuss about linear stability analysis for our new model by fluctuating all fields.
\end{abstract}

\maketitle

\section{Introduction}\label{sc:intr}
In recent years,
manifolds living in a higher dimensional space-time well known as \emph{D-branes} have fascinated numerous physicists. 
They have been derived from topological soliton solutions in string theory {\cite{Polchinski:1995mt}},
which arise naturally in ten-dimensional supergravity or string/M-theory
{\cite{Horava:1995qa,Horava:1996ma,Lukas:1998yy,Polchinski:1998rq,Polchinski:1998rr}}.  
It is an old idea that space-time may have more than four dimensions and the extra ones are unobservable for ordinary energy scales
{\cite{Kaluza:1921tu,Klein:1926tv,Rubakov:1983bb,Akama:1982jy,Visser:1985qm}}.
There are generally three different possibilities of extra-dimensional scenario:
extra-dimensional space is compact {\cite{Overduin:1998pn}},
non-compact but has a finite size {\cite{Antoniadis:1998ig,ArkaniHamed:1998rs,Randall:1999ee}},
and non-compact with an infinite size {\cite{Randall:1999vf}}. 
In particular,
Ref.{\cite{Randall:1999vf}} leads to localized gravity around a 3-brane with non-zero tension
in anti-de Sitter (AdS) space though the extra-dimensional space with an infinite size.

The brane proposed in Refs.{\cite{Randall:1999ee,Randall:1999vf}} and its generalizations,
\emph{i.e.}, models of Randall-Sundrum (RS) type,
are essentially static point-like external sources in the extra dimensions.
On the other hand,
an increasing interest recently has focused on study of gravitating thick (or fat) defects embedded in higher dimensional space-time
with codimension one or more 
{\cite{Csaki:2000fc,Kanno:2004nr,Gherghetta:2000jf,Gherghetta:2000qi,Gregory:1999gv,RandjbarDaemi:2000ft,Oda:2001ss}}. 
Our main proposal is to construct a new 3-brane which is described by special classical solutions,
\emph{i.e.}, topological solitons of a field-theoretical Lagrangian.
This approach is inspired by the D-branes where the solutions are constructed by topological solitons in string theory.
In that sense,
the solitonic branes may be more natural rather than another descriptions by the delta-function like potential.

3-branes in relevant previous works are constructed by the classical solutions in gauge theory;
such as kink in five dimensions (5D), Abelian-Higgs vortex in 6D, and t'Hooft-Polyakov monopole in 7D
{\cite{Antunes:2002hn,Giovannini:2006ye,Giovannini:2001hh,Brihaye:2006pi,Roessl:2002rv}}. 
In particular,
codimension-2 braneworld models constructed from Abelian-Higgs vortex in 6D are studied for a only case of flat 3-brane but also curved 3-brane
{\cite{Brihaye:2006pi}}.  
On the other hand,
our 3-brane is written by a Maxwell gauged $\mathbb{C}P^1$ model whose origin is $O(3)$ sigma model in (2+1)-dimensional field theory
{\cite{Manton:2004tk,Rajaraman:1982is,Arthur:1996uu,Piette:1994xf,Tchrakian:1995ji,Keeffe:1998kn}}.
$O(3)$ sigma model appears in various aspects of physics,
and has richer topological classes than Abelian-Higgs vortex; lump, gauged vortex, baby skyrmion, and instanton. 
The $O(3)$ sigma model can possess finite energy soliton solutions by adding some additional terms or gauges into the Lagrangian.
The solitons in gauge theory are strongly restricted by the configurations of their gauge groups;
on the contrary,
our framework is flexible and thus can be easily applied for various background geometry.

Another aim of the present paper is to analyze linear stability of our new solutions by fluctuating all fields
{\cite{Misner:1974qy,Weinberg:1972js,Mukhanov:1990me,Carmeli:2001ay}}.
Study of a linear stability and the second order correction of gravity and coupled fields for models of RS type have been done by numerous authors
{\cite{Flachi:2003bb,Garriga:1999yh,Garriga:2001ar,Goldberger:1999uk,Kudoh:2001wb,Sasaki:1999mi,Shiromizu:1999wj,Tanaka:2000zv}}.
Also,
analysis for gravitating thick defects embedded in higher dimensions are found in the literature;
for 5D {\cite{Giovannini:2001xg,Giovannini:2002jf,Giovannini:2006ye}}, and 
for 6D {\cite{Giovannini:2002mk,Giovannini:2002sb,Peter:2003zg,RandjbarDaemi:2002pq}}.
(Note that the model in Ref.{\cite{Giovannini:2006ye}} is constructed by gravitating multidefects in five dimensions).
The studies for thick defects, however, are {\it works in progress}
since the topological defects used in the literature are quite complicated structures.
More seriously,
stabilized mechanisms for the models of RS type, like ``moduli stabilization'', have not been found.
Analyzing the linear stability of our obtained solutions is thus worthwhile to tackle.

The plan of this paper is as follows.
In Sec.{\ref{sc:setup}} we build up the model and the resulting equations of motion. 
In Sec.{\ref{sc:bc}} we describe the boundary conditions for the matter fields and the warp factors.
In Sec.{\ref{sc:as}} we mainly discuss about the asymptotic behavior of the solutions of the model. 
In Sec.{\ref{sc:ns}} we introduce the methods for solving our boundary-value problem.
Some typical results localizing gravity around the 3-brane are shown in this section.
In Sec.{\ref{sc:lgp}} we give a detailed analysis for the stability of our branes for linearized gravitational perturbations.
Finally, in Sec.{\ref{sc:cncl}}  some conclusions and summary of this paper are drawn.

\section{Six-dimensional Model and Field Equations}\label{sc:setup}
In this section we shall construct Maxwell gauged $\mathbb{C}P^{1}$ model combining with general relativity in six dimensions.
As in the previous works for braneworld scenarios, 
an action of the models is written as a coupled system with gravity and solitons.
The total action of our six-dimensional model is written as
\begin{equation}
\mathcal{S} = \mathcal{S}_{\mathrm{brane}} + \mathcal{S}_{\mathrm{grav}},
\notag
\end{equation}
where $\mathcal{S}_{\mathrm{brane}}$ is Maxwell gauged $\mathbb{C}P^{1}$ model action and
$\mathcal{S}_{\mathrm{grav}}$ is the six-dimensional generalization of Einstein-Hilbert gravity.
The explicit form of Einstein-Hilbert gravity is given by
\begin{equation}
\mathcal{S}_{\mathrm{grav}} = {\int}d^{6}x \sqrt{-\mathcal{G}} \left( \frac{1}{2\chi}R - \Lambda_{b} \right),
\label{eq:grav_action}
\end{equation}
where $R$ is the six-dimensional Ricci scalar,
$\Lambda_{b}$ is the bulk cosmological constant, $\chi = 8{\pi}G_{6} = 8{\pi}/M_{6}^{4}$ and $M_{6}$ denotes the six-dimensional Planck mass.
On the other hand,
the brane action
\footnote{The conventions for indices of the present paper are the following:
the capital Latin indices run about from 0 to 5,
the Greek indices run about from 0 to 3,
and the small Latin indices which denote the complex projection ($\mathbb{C}P$) field,
therefore the small Latin indices run about from 1 to 2.} is of the form
\begin{multline}
\mathcal{S}_{\mathrm{brane}} = \int d^{6}x \sqrt{-\mathcal{G}}
\Biggl\{
- \frac{1}{4g}F_{MN}F^{MN}\\
- k(\mathcal{D}_{M}z^{a})^{*}\mathcal{D}^{M}z^{a}
- \mu\left[ 1 - z^{a*}\left(\sigma^{3}\right)^{ab}z^{b} \right]
\Biggr\},
\label{eq:brane_action}
\end{multline}
where $\sigma^{3}$ is the third component of the Pauli matrix.
The action (\ref{eq:brane_action}) is strongly motivated by the (2+1)-dimensional field-theoretical model {\cite{Tchrakian:1995ji,Arthur:1996uu}}. 
The coupling constants $g, k, \mu$ describe those of the field strength of the gauge field,
the kinetic term of $\mathbb{C}P^{1}$ and the potential term.
Their dimensions are $[M^{-2}], [M^4],[M^6]$ in the n.u.($M$ denotes a unit of mass), respectively.
The gauge covariant derivative $\mathcal{D}_{M}$ in Eq.(\ref{eq:brane_action}) is 
defined in terms of the $U(1)$ gauge field $A_{M}$ such as 
\begin{equation}
\mathcal{D}_{M} = \partial_{M} + iA_{M}\,.
\notag
\end{equation}
In standard $\mathbb{C}P^{1}$ model,
the $A_{M}$ is not an independent field but is the composite field connection
in terms of $\mathbb{C}P^{1}$ field defined as $\Tilde{A}_{M} = iz^{a*}\partial_{M}z^{a}$.
The composite field connection substantially works as a $U(1)$ gauge field {\cite{Rajaraman:1982is}}.
Thus we replace the composite field connection with the $U(1)$ gauge connection $A_{M}$
and treat as the independent field variable {\cite{Tchrakian:1995ji}}.
It should be noted that our model with this replacement is not to be completely equivalent to $O(3)$ sigma model.  
  
Formally by varying the actions (\ref{eq:grav_action}) and (\ref{eq:brane_action}) with respect to the field variables,
one can obtain the classical equations of motion
\begin{align}
&R_{MN} - \frac{1}{2}\mathcal{G}_{MN}R = \chi \left( T_{MN} - \Lambda_{b}\mathcal{G}_{MN} \right), \\ 
&\nabla_{M}\left(\mathcal{D}^{M}z^{a}\right) + iA_{M}\mathcal{D}^{M}z^{a} + \frac{\mu}{k}\sigma_{3}^{ab}z^{b} = 0, \\
&\nabla_{M}F^{MN} = 8kgA^{N}\left|\bm{z}\right|^{2} + i4kg\left(z^{a}\partial^{N}z^{a*} - z^{a*}\partial^{N}z^{a}\right),
\label{eqs:eqs}
\end{align}
where $\nabla_{M}$ is the covariant derivative with respect to the metric tensor and
\begin{align}
T_{MN}
&= - 2\frac{\delta \mathcal{L}_{\mathrm{brane}}}{\delta \mathcal{G}^{MN}}
+ \mathcal{G}_{MN}\mathcal{L}_{\mathrm{brane}}
\notag\\
&= 2k\left(\mathcal{D}_{M}z^{a}\right)^{*}\left(\mathcal{D}_{N}z^{a}\right)
+ \frac{1}{g}F_{MA}F_{N}^{A}
+ \mathcal{G}_{MN}\mathcal{L}_{\mathrm{brane}}
\label{eq:em_tensor}
\end{align}
is the six-dimensional energy-momentum tensor.

At the present paper, 
we would like to consider about a warped six-dimensional space-time with axially symmetric two extra dimensions. 
The ansatz is imposed on the metric tensor
\begin{align}
ds^{2} &= \mathcal{G}_{MN}dx^{M}dx^{N}
\notag\\
&= M^{2}(r)\eta_{\mu\nu}dx^{\mu}dx^{\nu}+dr^{2}+L^{2}(r)d\theta^{2},
\label{eq:metric}
\end{align}
where $r$ and $\theta$ are, respectively, the bulk radius and the bulk angle,
$\eta_{\mu\nu}$ denotes four-dimensional Minkowski metric tensor and
the function $M(r), L(r)$ are often called \emph{warp factors} in brane world scenarios.
For the matter fields,
we explore the solutions for the $\mathbb{C}P^{1}$ doublet field and the $U(1)$ gauge field of the form:
\begin{align}
&\bm{z} = (z_{1}, z_{2})^{T} = (\cos{\frac{f(r)}{2}}e^{-in\theta}, \sin{\frac{f(r)}{2}})^{T},
\notag\\
&A_{\mu} = 0,\;\;\; A_{r} = 0,\;\;\; A_{\theta} = n - a(r),
\label{eqs:matter_ansatz}
\end{align}
where $n$ is the winding number of the gauge field and the $\mathbb{C}P^{1}$ doublet $\bm{z}$ satisfies the constraint $\bm{z}^{\dag}\bm{z} = 1$.

The classical equations of motion of the present system are then
\begin{align}
&\Tilde{f}^{\prime\prime} + ( 4m + \ell )\Tilde{f}^{\prime} - 2\left( \frac{v_{f}}{\mathcal{L}^{2}} + \gamma\sin{\Tilde{f}} \right) = 0,
\label{eq:scalar}\\
&\Tilde{a}^{\prime\prime} + ( 4m - \ell )\Tilde{a}^{\prime} - v_{a} = 0,
\label{eq:gauge}\\
&\ell^{\prime} + 3m^{\prime} + \ell^{2} + 6m^{2} + 3m\ell = \alpha\left( \tau_{0} - \beta \right),
\label{eq:eeq_4d}\\
&4m^{\prime} + 10m^{2} = \alpha\left( \tau_{\theta} - \beta \right),
\label{eq:eeq_theta}\\
&4m\ell + 6m^{2} = \alpha\left( \tau_{r} - \beta \right),
\label{eq:eeq_r}
\end{align}
where $\alpha := {\chi}k$ is the dimensionless gravitational coupling constant,
$\beta := \Lambda_{b}/k^{2}g$ is the dimensionless bulk cosmological constant,
and $\gamma := \mu/k^{2}g$ is the dimensionless coupling constant.
Besides the dimensionless coordinate
\begin{equation}
\rho := \sqrt{kg}r
\notag
\end{equation}
is introduced and the prime in the equations denotes the derivative with respect to this dimensionless coordinate.
The matter fields $f(r)$, $a(r)$ and the warp factors $M(r)$, $L(r)$ are rewritten in terms of the new coordinate,
that is, 
\begin{align}
&\Tilde{f}(\rho) := f(r), &&\Tilde{a}(\rho) := a(r),
\notag\\
&\mathcal{M}(\rho) := M(r), &&\mathcal{L}(\rho) := \sqrt{kg}L(r).
\notag
\end{align}
The functions $m(\rho)$ and $\ell(\rho)$ are defined by the above functions as
\begin{equation}
m(\rho) := \frac{d \ln \mathcal{M}(\rho)}{d\rho},\;\;\; \ell(\rho) := \frac{d \ln \mathcal{L}(\rho)}{d\rho}
\label{eqs:dimensionless_CH_trans}
\end{equation}
which seem to have some similarity with the Cole-Hopf transformation appearing in the integrable theory.
The components of the energy-momentum tensor $T_{M}^{N}$ go to be dimensionless ones
\begin{align}
\tau_{0}(\rho)
&:= \frac{T_{0}^{0}}{k} = - \frac{1}{4}\Tilde{f}^{\prime2} - \frac{\Tilde{a}^{\prime2}}{2\mathcal{L}^{2}} - \frac{v}{\mathcal{L}^{2}} - \gamma(1 - \cos{\Tilde{f}}),
\notag\\
\tau_{r}(\rho)
&:= \frac{T_{r}^{r}}{k} = \frac{1}{4}\Tilde{f}^{\prime2} + \frac{\Tilde{a}^{\prime2}}{2\mathcal{L}^{2}} - \frac{v}{\mathcal{L}^{2}} - \gamma(1 - \cos{\Tilde{f}}),
\notag\\
\tau_{\theta}(\rho)
&:= \frac{T_{\theta}^{\theta}}{k}
= - \frac{1}{4}\Tilde{f}^{\prime2} + \frac{\Tilde{a}^{\prime2}}{2\mathcal{L}^{2}} + \frac{v}{\mathcal{L}^{2}} - \gamma(1 - \cos{\Tilde{f}}),
\label{eqs:em_tensor}
\end{align}
where the dimensionless quantities $v,v_{a},v_{f}$ have been defined as
\begin{align}
&v := a^{2} + n(n - 2a)\sin^{2}\frac{f}{2}, 
\notag\\
&v_{a} := \frac{\partial{v}}{\partial{a}} =  2\left( a - n\sin^{2}{\frac{f}{2}} \right), 
\notag\\
&v_{f} := \frac{\partial{v}}{\partial{f}} = \frac{n(n - 2a)}{2}\sin{f}. 
\label{eqs:d_less_vs}
\end{align}
  
Before beginning our analysis,
let us consider basic properties of the field equations (\ref{eq:scalar})-(\ref{eq:eeq_r}).
Clearly Eqs.(\ref{eq:scalar}) and (\ref{eq:gauge}) are the dynamical equations
because they are second order differential equations of the fields $\Tilde{f}(\rho), \Tilde{a}(\rho)$.
Since Eqs.(\ref{eq:eeq_4d}) and (\ref{eq:eeq_theta}) contain $\ell^{\prime}, m^{\prime}$,
\emph{i.e.}, the second derivative of the warp factors,
these equations are the dynamical equations, too. 
Eq.(\ref{eq:eeq_r}) is comprised of the first derivatives only.
Thus it means that the equation works as a constraint equation for the dynamical fields $\Tilde{f}(\rho), \Tilde{a}(\rho), \mathcal{M}(\rho), \mathcal{L}(\rho)$.
As a result,
we must treat a numerical problem of a series of four dynamical equations with one constraint equation.

\section{Boundary Conditions}\label{sc:bc}

\subsection{Soliton}
Existence of topological soliton solutions is inferred from Derrick's scaling argument {\cite{Derrick:1964ww}}
in which, if soliton exists,
a stationary point of the energy in the field configuration should be stationary against all variations including spatial rescaling.
Also,
the soliton solutions always have the lower energy bound which is defined by their topology.
For the $U(1)$ gauged $\mathbb{C}P^{1}$ model with Maxwell or Chern-Simon term,
such existence proof was confirmed numerically {\cite{Tchrakian:1995ji,Piette:1994xf,Arthur:1996uu}}.
Besides the original Maxwell gauged $\mathbb{C}P^{1}$ model in (2+1) dimensions can have topological soliton solutions {\cite{Tchrakian:1995ji}}.
These analysis clearly indicate that the solitons can be stabilized without any higher order terms,
like the \emph{Skyrme term} in (3+1) dimensions {\cite{Manton:2004tk}}.

If we apply these models to the six-dimensional space-time,
both the $\mathbb{C}P^1$ field and the $U(1)$ gauge field should go to zero at infinity and be regular at the origin in the extra-dimensional space,
which exactly agree with the topological requirements of the model. 
These conditions are
\begin{align}
&\Tilde{f}(0) = \pi, &&\lim_{\rho \to \infty}\Tilde{f}(\rho) = 0,
\notag\\
&\Tilde{a}(0) = n, &&\lim_{\rho \to \infty}\Tilde{a}(\rho) = 0.
\label{eqs:matter_bc}
\end{align}
As is well known that $\mathbb{C}P^{N}$ model has a close relation to $O(3)$ sigma model.
For the case of $N=1$,
one can easily obtain the $O(3)$ sigma model by using the transformation
\begin{equation}
\phi^{\Tilde{a}} := z^{*a}\left(\sigma^{\Tilde{a}}\right)^{ab}z^{b},
\notag
\end{equation}
where $\phi^{\Tilde{a}}, \Tilde{a}=1, 2, 3$ are scalar triplet fields,
$z^{a}$ are doublet fields,
and $\sigma^{\Tilde{a}}$ are the three Pauli matrices. 
In the sense,
the original $\mathbb{C}P^{1}$ model is essentially $O(3)$ sigma model when it is written in terms of the $\phi^{\Tilde{a}}$.
Moreover, in the view of $O(3)$ sigma model,
the topological boundary condition is interpreted as the south pole configuration at the origin and the north pole configuration at infinity.
The details of above discussion can be seen, \emph{e.g.}, in Ref.{\cite{Rajaraman:1982is}}.
As mentioned before,
our model however is not equivalent to $O(3)$ sigma model completely.

\subsection{Geometry}
Regular geometry at the origin defines the boundary conditions for the warp factors.
An effective action derived from six-dimensional gravity action generally has two kinds of singularity at the origin,
which are called conical and curvature singularity problem (\emph{e.g.}, in Ref.{\cite{Papantonopoulos:2006uj}}).
The boundary conditions for geometry are introduced in order to exclude these serious difficulties in the six-dimensional model.
In this paper they are given by
\begin{equation}
\mathcal{M}^{\prime}(0) = 0, \;\;\; \mathcal{L}(0) = 0, \;\;\; \mathcal{L}^{\prime}(0) = 1.
\label{eq:reg}
\end{equation}
One simply fixes $\mathcal{M}(0) = 1$ since the value of $\mathcal{M}(0)$ is an arbitrary constant.
The boundary conditions for $\mathcal{L}^{\prime}$ and $\mathcal{M}^{\prime}$
correspond to the conical and the curvature singularity problem at the origin, respectively.
Also,
we concentrate on the problem of the regular geometry without the conical singularity
so that we employ the boundary condition $\mathcal{L}^{\prime}(0) = 1$.

In order to solve the equations of motion (\ref{eq:scalar})-(\ref{eq:eeq_r}) 
by an analytical method, the boundary conditions (\ref{eqs:matter_bc}) and (\ref{eq:reg}) are sufficient.
However,
Eqs.(\ref{eq:reg}) do not tell anything about the asymptotic behavior of the metric tensor at large $\rho$ and
thus a possibility of the gravity localization around the 3-brane remains unknown.
The requirement of the gravity localization is equivalent to a finiteness of the four-dimensional Planck mass $M_{\mathrm{Planck}}$,
\emph{i.e.}, an inequality
\begin{equation}
M_{\mathrm{Planck}}^{2} \sim 2\pi M_{6}^{4} \int d\rho \mathcal{M}^{2}(\rho)\mathcal{L}(\rho) < \infty
\label{eq:planckmass}
\end{equation}
must be satisfied.
Notice that solutions respecting the boundary conditions do not often satisfy the inequality (\ref{eq:planckmass}). 
Since the inequality requires a fine-tuning of parameters for realizing the gravity localization,
practically it works as the fifth boundary condition of the model.
As we shall see below,
imposing it on the model and considering about a empty space-time with only the bulk cosmological constant,
we can find the information of geometry far from the origin.

\subsection{Vacuum solution of warp factors}\label{sbsc:vacuum}
If a gravitational source in certain brane model is constructed by a \emph{local} topological defect,
the all source terms in the Einstein equation will vanish for the region of $\rho\to\infty$.
Namely an asymptotic behavior of geometry at infinity,
which can not be determined by the boundary conditions,
obeys vacuum solutions of the sourceless Einstein equation with the bulk cosmological constant.

Combining the components of Einstein equation (\ref{eq:eeq_4d})-(\ref{eq:eeq_r}),
one can easily find equations for vacuum solutions,
given by
\begin{equation}
m^{\prime} + \frac{5}{2}m^{2} + \frac{\alpha\beta}{4} = 0,\;\;\;
\ell = - \frac{\alpha\beta}{4m} - \frac{3m}{2}.
\notag
\end{equation}
If the cosmological constant $\beta$ is negative,
then one analytically obtains the solutions
\begin{align}
&\mathcal{M}(\rho) = \mathcal{M}_{0}e^{-c\rho}|1+\epsilon e^{5c\rho}|^{2/5},
\notag\\
&\mathcal{L}(\rho) = \mathcal{L}_{0}e^{-c\rho}|\epsilon e^{5c\rho} - 1|\cdot|1 + \epsilon e^{5c\rho}|^{-3/5},
\label{eqs:gv_solutions}
\end{align}
where $\mathcal{M}_{0}, \mathcal{L}_{0}, \epsilon$ are arbitrary integral constants
and the coefficient
\begin{equation}
c:=\sqrt{- \alpha\beta/10}>0
\label{eq:parameter_c}
\end{equation}
is a function of the model parameters.
On the other hand,
the case for the positive $\beta$ was investigated in Ref.{\cite{Rubakov:1983bz}}.
Therefore we shall concentrate the case of the negative bulk cosmological constant $\beta$ at the present paper.
Both cases are acceptable in seven dimensions {\cite{Roessl:2002rv}}.

Furthermore,
in order to study the singularity structure of the geometry,
we investigate all curvature invariants for the metric tensor (\ref{eq:metric}).
Explicit forms of the curvature invariants are
\begin{align}
&\mathcal{R} := \frac{R}{kg}
= - 8\frac{\mathcal{M}^{\prime}\mathcal{L}^{\prime}}{\mathcal{M}\mathcal{L}}
- 12\frac{\mathcal{M}^{\prime2}}{\mathcal{M}^{2}}
- 2\frac{\mathcal{L}^{\prime\prime}}{\mathcal{L}}
- 8\frac{\mathcal{M}^{\prime\prime}}{\mathcal{M}},
\notag\\
&\mathcal{R}_{MN}\mathcal{R}^{MN} := \frac{R_{MN}R^{MN}}{(kg)^{2}}
\notag\\
&= 20\frac{\mathcal{M}^{\prime2}\mathcal{L}^{\prime2}}{\mathcal{M}^{2}\mathcal{L}^{2}}
+ 24\frac{\mathcal{M}^{\prime3}\mathcal{L}^{\prime}}{\mathcal{M}^{3}\mathcal{L}}
+ 36\frac{\mathcal{M}^{\prime4}}{\mathcal{M}^{4}}
+ 8\frac{\mathcal{M}^{\prime}\mathcal{L}^{\prime}\mathcal{L}^{\prime\prime}}{\mathcal{M}\mathcal{L}^{2}}
\notag\\
&+ 2\frac{\mathcal{L}^{\prime\prime2}}{\mathcal{L}^{2}}
+ 8\frac{\mathcal{M}^{\prime\prime}\mathcal{M}^{\prime}\mathcal{L}^{\prime}}{\mathcal{M}^{2}\mathcal{L}}
+ 24\frac{\mathcal{M}^{\prime\prime}\mathcal{M}^{\prime2}}{\mathcal{M}^{3}}
+ 8\frac{\mathcal{M}^{\prime\prime}\mathcal{L}^{\prime\prime}}{\mathcal{M}\mathcal{L}}
\notag\\
&+ 20\frac{\mathcal{M}^{\prime\prime2}}{\mathcal{M}^{2}},
\notag\\
&\mathcal{R}_{ABCD}\mathcal{R}^{ABCD} := \frac{R_{ABCD}R^{ABCD}}{(kg)^{2}}
\notag\\
&= 16\frac{\mathcal{M}^{\prime2}\mathcal{L}^{\prime2}}{\mathcal{M}^{2}\mathcal{L}^{2}}
+ 24\frac{\mathcal{M}^{\prime4}}{\mathcal{M}^{4}}
+ 16\frac{\mathcal{M}^{\prime\prime2}}{\mathcal{M}^{2}}
+ 4\frac{\mathcal{L}^{\prime\prime2}}{\mathcal{L}^{2}},
\notag\\
&\mathcal{C}_{ABCD}\mathcal{C}^{ABCD} := \frac{C_{ABCD}C^{ABCD}}{(kg)^{2}}
\notag\\
&= \frac{12\left[
\mathcal{M}^{\prime2}\mathcal{L}
+ \mathcal{M}^{2}\mathcal{L}^{\prime\prime}
- \mathcal{M}\left( \mathcal{M}^{\prime}\mathcal{L}^{\prime} + \mathcal{M}^{\prime\prime}\mathcal{L} \right)
\right]^{2}}
{5\mathcal{M}^{4}\mathcal{L}^{2}}
\label{eqs:curv_inv}
\end{align}
where $\mathcal{R}, \mathcal{R}_{MN}, \mathcal{R}_{ABCD}, \mathcal{C}_{ABCD}$ are defined as dimensionless curvatures.
The curvatures were already defined, \emph{e.g.}, in Ref.{\cite{Weinberg:1972js}}.

Inserting Eqs.(\ref{eqs:gv_solutions}) into Eqs.(\ref{eqs:curv_inv}),
we study the geometry at infinity.
The scalar and the Ricci curvature invariants
\begin{equation}
\mathcal{R} = - 30c^{2},\;\;\;
\mathcal{R}_{MN}\mathcal{R}^{MN} = 150c^{4}
\label{eqs:scalar&ricci}
\end{equation}
are simply constants for any $\epsilon$,
whereas the Riemann and the Weyl curvature invariants 
\begin{align}
\mathcal{R}_{ABCD}\mathcal{R}^{ABCD} = 60c^{4}F(\epsilon),
\notag\\
\mathcal{C}_{ABCD}\mathcal{C}^{ABCD} = \frac{3840c^{4}\epsilon^{2}e^{10c\rho}}{\left( 1 + \epsilon e^{5c\rho} \right)^{4}},
\label{eqs:riemann&weyl}
\end{align}
are functions of $\epsilon$,
where
\begin{equation}
F(\epsilon) :=
\frac{1 + 4\epsilon e^{5c\rho} + 70\epsilon^{2}e^{10c\rho} + 4\epsilon^{3}e^{15c\rho} + \epsilon^{4}e^{20c\rho}}
{\left( 1 + \epsilon e^{5c\rho} \right)^{4}}.
\notag
\end{equation}
Thus we find that
the scalar and the Ricci curvature invariants are always constant and the regularity of the Riemann and the Weyl curvature invariants depend on $\epsilon$,
for the above vacuum solutions.

Next,
in order to study whether the gravity can be localized and the geometry can be regularized on those solutions,
we must analyze about the integral constant $\epsilon$.

For $\epsilon = 0$, the solutions become
\begin{equation}
m(\rho) = \ell(\rho) = -c
\label{eq:bulk_solution_ml}
\end{equation}
and thus the warp factors
\begin{equation}
\mathcal{M}(\rho) = \mathcal{M}_{0}e^{-c\rho},\;\;\;\mathcal{L}(\rho) = \mathcal{L}_{0}e^{-c\rho}
\label{eq:bulk_solution}
\end{equation}
are exponentially decreasing.
In the present case the Riemann and the Weyl curvature invariants are also constants similar to the others.
Also each of the curvature invariants is equivalent to the one
calculated from Riemann tensor for AdS$_{N}$ space in Ref.{\cite{Satoukodama:2000iw}}.
For AdS$_{N}$ space,
each of the curvature tensors is represented in terms of the Gaussian curvature $K$,
which are given by
\begin{align}
&R = N\left( N - 1 \right)K,\;\;\; R_{AB} = \left( N - 1 \right)K\mathcal{G}_{AB},
\notag\\
&R_{ABCD} = K\left( \mathcal{G}_{AC}\mathcal{G}_{BD} - \mathcal{G}_{AD}\mathcal{G}_{BC} \right).
\notag
\end{align}
Since one can easily find $K = -c^{2}$ from $\mathcal{R}$ in Eqs.(\ref{eqs:scalar&ricci}) in our six-dimensional model (namely $N = 6$),
the other curvature invariants for AdS$_{6}$ can be also obtained simply. 
Therefore the choice of $\epsilon$ is desired one for the finiteness of the four-dimensional Planck mass (\ref{eq:planckmass});
since the asymptotic solutions can lead to a smooth AdS$_{6}$ geometry far from the vortex string core
and to localize gravity around the vortex string
\footnote{
Other cases of $\epsilon$ are as follows
(see also Ref.{\cite{Giovannini:2001hh}}).
If $\epsilon > 0$ or $\epsilon \leqslant -1$, 
we obtain exponentially growing solutions.
For this case Eq.(\ref{eq:planckmass}) diverges and therefore gravity can not be localized.
(The case $\epsilon = -1$ is somewhat specific since a singularity of Kasner type is developed in the origin.)
If $-1 < \epsilon < 0$,
the geometry has a singular point $\rho_{0}$ and $\rho < \rho_{0}$ should be required.
In spite of the fact,
the integral (\ref{eq:planckmass}) is finite.
Thus this case has a possibility of localizing gravity if the singularity at $\rho_{0}$ is resolved.
Finally,
let us note that if the bulk cosmological constant $\Lambda_{b}$ is zero,
solutions have a power-law behavior belonging to Kasner class.
These solutions leave open only two possibilities but can not lead to localization of gravity.}.

The integral constants $\epsilon, \mathcal{M}_{0}, \mathcal{L}_{0}$ have not been fixed for the above discussion.
However,
if a string-like defect is placed at the origin $\rho = 0$,
the constants are no longer arbitrary and become functions of the model parameters $\alpha, \beta, \gamma$, 
namely, $\epsilon = \epsilon(\alpha, \beta, \gamma)$, and so on.
The regular geometry is achieved together with the gravity localization if the parameters lie on the surface $\epsilon(\alpha, \beta, \gamma) = 0$.
Therefore we shall find numerical solutions with $\epsilon(\alpha, \beta, \gamma) = 0$.

\section{Asymptotic solutions}\label{sc:as}
In order to solve Eqs.(\ref{eq:scalar})-(\ref{eq:eeq_r}) numerically,
asymptotic behaviors of the warp factors, scalar, and gauge fields in the vicinity of the origin
as well as at large distance of the core are mandatory {\cite{William:1995nr,Keller:1992tb,Stoer:2002na}}.
They are obtained by expanding the functions around the origin and approximating the equations at infinity.
In this section,
we also study about relations for a string tension since they may give useful informations to find the proper solutions.

\subsection{At the origin}
To examine behaviors at the origin,
we start by expanding the warp factors together with the scalar and gauge fields as power series in $\rho$.
Here we consider the case of $n = 1$.
(For other $n$'s one can estimate in a quite similar fashion.) 
Inserting the power series into the equations of motion (\ref{eq:scalar})-(\ref{eq:eeq_r})
and requiring that the expanded equations obey the boundary conditions (\ref{eqs:matter_bc}) and (\ref{eq:reg}) for a limit $\rho \to 0$,
then one can get the asymptotic solutions 
\begin{align}
&\Tilde{f}(\rho) \simeq \pi + \mathcal{A}\rho,
\notag\\
&\Tilde{a}(\rho) \simeq 1 + \mathcal{B}\rho^{2},
\notag\\
&\mathcal{M}(\rho) \simeq 1 + \frac{\alpha}{8}\left( - \beta - 2\gamma + 2\mathcal{B}^{2} \right)\rho^{2},
\notag\\
&\mathcal{L}(\rho) \simeq \rho - \frac{\alpha}{12}\left[ - \beta - 2\gamma + \mathcal{A}^{2} + 10\mathcal{B}^{2} \right]\rho^{3},
\label{eqs:asymp_origin}
\end{align}
where the coefficients $\mathcal{A}$ and $\mathcal{B}$ are two arbitrary constants.
They can not clearly be determined by only locally analyzing the equations of motion.
We thus need informations of the results of numerical integration.
Practically,
the constants are used to realize the boundary conditions at infinity.

Furthermore,
inserting the asymptotic solutions (\ref{eqs:asymp_origin}) into Eqs.(\ref{eqs:em_tensor}) and (\ref{eqs:curv_inv}),
we find the asymptotic behaviors of the energy-momentum tensor and the curvature invariants in the vicinity of the origin.
We obtain the asymptotic forms around the origin
\begin{align}
&\tau_{0}(\rho) \simeq - 2\gamma - \frac{\mathcal{A}^{2}}{2} - 2\mathcal{B}^{2} + \mathcal{O}(\rho),
\notag\\
&\tau_{r}(\rho) \simeq - 2\gamma + 2\mathcal{B}^{2} + \mathcal{O}(\rho),
\notag\\
&\tau_{\theta}(\rho) \simeq - 2\gamma + 2\mathcal{B}^{2} + \mathcal{O}(\rho)
\end{align}
for the components of energy-momentum tensor and
\begin{align}
&\mathcal{R}
\simeq \alpha\left[
3\left( \beta + 2\gamma \right)
+ \mathcal{A}^{2}
+ 2\mathcal{B}^{2}
\right] + \mathcal{O}(\rho), 
\notag\\
&\mathcal{R}_{AB}\mathcal{R}^{AB}
\simeq \frac{\alpha^{2}}{2}\Bigl[
3\left( \beta+2\gamma \right)^{2}
+ 2\left( \beta+2\gamma \right)\left( \mathcal{A}^{2} + 2\mathcal{B}^{2} \right)
\notag\\
&\phantom{\mathcal{R}_{AB}\mathcal{R}^{AB}\simeq}
+ \mathcal{A}^{4}
+ 12\mathcal{A}^{2}\mathcal{B}^{2}
+ 44\mathcal{B}^{4}
\Bigr] + \mathcal{O}(\rho),
\notag\\
&\mathcal{R}_{ABCD}\mathcal{R}^{ABCD} 
\simeq \alpha^{2}\Bigl[
2\left( \beta + 2\gamma - 2\mathcal{B}^{2} \right)^{2}
\notag\\
&\phantom{\mathcal{R}_{ABCD}\mathcal{R}^{ABCD} \simeq}
+ \left( \beta + 2\gamma - \mathcal{A}^{2} - 10\mathcal{B}^{2} \right)^{2}
\Bigr] + \mathcal{O}(\rho),
\notag\\
&\mathcal{C}_{ABCD}\mathcal{C}^{ABCD}
\simeq \frac{3}{5}\alpha^{2}\left[
\mathcal{A}^{2}
+ 12\mathcal{B}^{2}
- 2\left( \beta + 2\gamma \right)
\right]^{2}
\notag\\
&\phantom{\mathcal{C}_{ABCD}\mathcal{C}^{ABCD}\simeq}
+ \mathcal{O}(\rho)
\end{align}
for the curvature invariants, respectively.

\subsection{At infinity}
Next we will consider asymptotic solutions for the two matter fields at large $\rho$.
Let us assume that geometry is regular at infinity as the case $\epsilon = 0$.
Then asymptotic solutions of the warp factors are determined as Eqs.(\ref{eq:bulk_solution_ml})
\footnote{The same geometry can be also realized in the case of $\epsilon > 0$ or $\epsilon < -1$,
however both cases do clearly not realize the inequality (\ref{eq:planckmass}) for gravity localization.
Thus we exclude these possibilities in subsequent consideration.}.

From the boundary conditions for the matter fields (\ref{eqs:matter_bc}),
asymptotic forms of them can be expressed by
\begin{equation}
\Tilde{f}(\rho) = \delta f(\rho),\;\;\; \Tilde{a}(\rho) = \delta a(\rho).
\notag
\end{equation}
Inserting the asymptotic forms and Eq.(\ref{eq:bulk_solution_ml}) into the equation of motion for the gauge field (\ref{eq:gauge}),
one easily obtains
\begin{equation}
(\delta a)^{\prime\prime} - 3c(\delta a)^{\prime} - 2(\delta a) = 0.
\notag
\end{equation}
For $\rho \gg 1$ the solution $\Tilde{a}(\rho)$ can be described approximately by
\begin{equation}
\delta a(\rho) \sim e^{-q\rho},\;\;\; q = - \frac{3c}{2}\left( 1 \pm \sqrt{1 + \frac{8}{9c^{2}}} \right).
\label{eq:asymptotic_a}
\end{equation}  
Similarly,
the linearized $\mathbb{C}P^{1}$ field equation yields
\begin{equation}
(\delta f)^{\prime\prime} - 5c(\delta f)^{\prime} - 2 \gamma(\delta f) = 0
\notag
\end{equation}
which leads to the solution for the $\mathbb{C}P^{1}$ field
\begin{equation}
\delta f(\rho) \sim e^{-p\rho},\;\;\; p = - \frac{5c}{2}\left( 1 \pm \sqrt{1 + \frac{8\gamma}{25c^{2}}} \right).
\label{eq:asymptotic_f}
\end{equation}
In the asymptotic solutions (\ref{eq:asymptotic_a}) and (\ref{eq:asymptotic_f}),
a minus sign should be chosen in order to obtain exponentially decreasing behaviors.
Since the asymptotic solutions must also satisfy the boundary conditions (\ref{eqs:matter_bc}) at infinity,
the other exponentially growing solutions can never be allowed.

For the completeness of our analysis,
let us study the accuracy of our approximations more in detail.
The equation of motion for the $\mathbb{C}P^{1}$ field (\ref{eq:scalar}) and the components of the energy-momentum tensor (\ref{eqs:em_tensor})
contain the terms of order $\mathcal{O}(1/\mathcal{L}^{2})$,
which exponentially diverge at large $\rho$.
The linearization of Eq.(\ref{eq:scalar}) is justified if and only if $q > 2c$,
\emph{i.e.}, $2 > \alpha|\beta|$ which can be seen by easy calculations.
Also, as we shall see in Sec.{\ref{sbsc:vacuum}},
in order to drop the components of the energy-momentum tensor rather than the warp factors,
we find that $p > c$ and $q > c$,
\emph{i.e.}, $10\gamma > 3\alpha|\beta|$ and $5 > \alpha|\beta|$ are required.
Thus, especially for strong gravity limit,
we need much thorough analyses for finding asymptotic solutions.

\subsection{Relationship for the string tension}
In four-dimensional cosmology,
vortex-like topological defects often called cosmic string have been extensively studied.
In particular, components of the string tension defined as
\begin{equation}
\mu_{M} := \int_{0}^{\infty}d\rho \mathcal{M}^{4}(\rho)\mathcal{L}(\rho)\tau_{M}(\rho)
\notag
\end{equation}
contain many useful informations for the geometry {\cite{Frolov:1989er}}.
In the study of Abelian vortex in six dimensions {\cite{Giovannini:2001hh}},
a constraint is easily found by a reduction which we shall demonstrate in the below.
Furthermore,
the constraint plays an essential role in the stability analysis of vector mode fluctuations {\cite{Giovannini:2002sb,Giovannini:2002mk}}.
In our model a similar relation can be found but the usefulness is rather problematic:
because it contains a functional of $\Tilde{f}(\rho), \Tilde{a}(\rho), \mathcal{M}(\rho), \mathcal{L}(\rho)$. 
In the case of 'tHooft-Polyakov monopole in seven dimensions {\cite{Roessl:2002rv}},
the situation is quite similar to ours.

Consider two specific linear combinations of Einstein equation (\ref{eq:eeq_4d})-(\ref{eq:eeq_r}),
namely
\begin{align}
&m^{\prime} + 4m^{2} + m\ell = \frac{\alpha}{4}(\tau_{\theta} + \tau_{r}) - \frac{\alpha\beta}{2},
\notag\\
&\ell^{\prime} + \ell^{2} + 4m\ell = \frac{\alpha}{4}(4\tau_{0} + \tau_{r} - 3\tau_{\theta}) - \frac{\alpha\beta}{2}.
\label{eqs:linear_com}
\end{align}
Integrating Eqs.(\ref{eqs:linear_com}) from zero to infinity,
one easily find the relations
\begin{align}
&\lim_{\rho_{c}\to\infty}\mathcal{M}^{3}(\rho_{c})\mathcal{M}_{\rho}(\rho_{c})\mathcal{L}(\rho_{c})
\notag\\
&= \frac{\alpha}{4}(\mu_{\theta} + \mu_{r}) - \frac{\alpha\beta}{2}\int_{0}^{\infty}\mathcal{M}^{4}\mathcal{L}d\rho,
\label{eq:tolman}\\
&\lim_{\rho_{c}\to\infty}\mathcal{M}^{4}(\rho_{c})\mathcal{L}_{\rho}(\rho_{c}) - 1
\notag\\
&= \frac{\alpha}{4}(4\mu_{0} + \mu_{r} - 3\mu_{\theta}) - \frac{\alpha\beta}{2}\int_{0}^{\infty}\mathcal{M}^{4}\mathcal{L}d\rho,
\label{eq:deficit}
\end{align}
where the both right-hand-sides of them have already been taken to the limit $\rho_{c}\to\infty$.
Let us note that we do not impose the localizing condition of gravity (\ref{eq:planckmass}) on the above formulations yet.
In the limit $\rho_{c} \to \infty$,
Eq.(\ref{eq:tolman}) is the six-dimensional analogue of the relation determining the Tolman mass
whereas Eq.(\ref{eq:deficit}) is the generalization of the relation giving the deficit angle.
Imposing the condition (\ref{eq:planckmass}) on Eqs.(\ref{eq:tolman}) and (\ref{eq:deficit}) and subtracting Eq.(\ref{eq:tolman}) from Eq.(\ref{eq:deficit}),
one obtains the following relation
\footnote{If $\epsilon \ne 0$,
one can use other asymptotic solutions of the metric obtained by previous discussion
for computation of the left-hand-sides of Eqs.(\ref{eq:tolman}) and (\ref{eq:deficit}).
Then this calculation gives a more general relation {\cite{Giovannini:2001hh}}.}
\begin{equation}
\mu_{0} - \mu_{\theta} = - \frac{1}{\alpha}
\label{eq:fine_tune}
\end{equation}
which must be hold where $\mu_{r}$ still remains undermined.
In fact we do not use $\mu_{r}$ now.
The condition (\ref{eq:fine_tune}) is, however, only necessary but not sufficient in order to have solutions leading to localized gravity.

In order to get some informations of solutions at the origin,
we directly integrate the left-hand-side of Eq.(\ref{eq:fine_tune}) and obtain
\begin{equation}
\mu_{0} - \mu_{\theta} = - \int_{0}^{\infty}d\rho\frac{\mathcal{M}^{4}}{\mathcal{L}}\left( 2v + \Tilde{a}^{\prime2} \right).
\label{eq:left_1}
\end{equation}
Using the equation of motion for the gauge field (\ref{eq:gauge}),
the following relation is obtained
\begin{equation}
\left( \frac{\mathcal{M}^{4}\Tilde{a}^{\prime}}{\mathcal{L}} \right)^{\prime}
= \frac{\mathcal{M}^{4}}{\mathcal{L}}v_{a}
= 2\frac{\mathcal{M}^{4}}{\mathcal{L}}\left( \Tilde{a} - n\sin^{2}{\frac{\Tilde{f}}{2}} \right),
\notag
\end{equation}
where Eqs.(\ref{eqs:d_less_vs}) are used at the last step above formulation.
Inserting back it into Eq.(\ref{eq:left_1}),
the integral in Eq.(\ref{eq:left_1}) can be rewritten as
\begin{align}
&\mu_{0} - \mu_{\theta}
\notag\\
= &- \int_{0}^{\infty}d\rho\frac{\mathcal{M}^{4}}{\mathcal{L}}
\left\{
2\left[ \Tilde{a}^{2} + n\left( n - 2\Tilde{a} \right)\sin^{2}\frac{\Tilde{f}}{2} \right] + \Tilde{a}^{\prime2}
\right\}
\notag \\
= &- \int_{0}^{\infty}d\rho
\left[
2\frac{\mathcal{M}^{4}}{\mathcal{L}}n\left( n - \Tilde{a} \right)\sin^{2}\frac{\Tilde{f}}{2}
+ \left( \frac{\mathcal{M}^{4}\Tilde{a}^{\prime}\Tilde{a}}{\mathcal{L}} \right)^{\prime}
\right]
\notag\\
= &- 2\varXi\left[ \Tilde{f}, \Tilde{a}, \mathcal{M}, \mathcal{L} \right]
- \left(
\left.\frac{\mathcal{M}^{4}\Tilde{a}^{\prime}\Tilde{a}}{\mathcal{L}}\right|_{\infty}
- \left.\frac{\mathcal{M}^{4}\Tilde{a}^{\prime}\Tilde{a}}{\mathcal{L}}\right|_{0}
\right),
\label{eq:left_2}
\end{align}
where for order we defined a functional of $\Tilde{f}, \Tilde{a}, \mathcal{M}, \mathcal{L}$ as
\begin{equation}
\varXi\left[ \Tilde{f}, \Tilde{a}, \mathcal{M}, \mathcal{L} \right]
:= \int_{0}^{\infty}d\rho\frac{\mathcal{M}^{4}}{\mathcal{L}}n\left( n - \Tilde{a} \right)\sin^{2}\frac{\Tilde{f}}{2}.
\notag
\end{equation}
For solutions the leading gravity localization,
the boundary term at infinity in Eq.(\ref{eq:left_2}) exactly vanishes.
Moreover,
using the boundary conditions at the origin and the fine-tuning relation (\ref{eq:fine_tune}),
we obtain
\begin{equation}
- \frac{1}{\alpha} = - 2\varXi\left[ \Tilde{f}, \Tilde{a}, \mathcal{M}, \mathcal{L} \right] + \left.\frac{\Tilde{a}^{\prime}}{\mathcal{L}}\right|_{0}.
\label{eq:fine_tune_2}
\end{equation}
According to the asymptotic solutions in the vicinity of the origin (\ref{eqs:asymp_origin}),
$\Tilde{a}(\rho)$ and $\mathcal{L}(\rho)$ are, respectively,
$\Tilde{a} \simeq 1 + \mathcal{B}\rho^{2}$ and $\mathcal{L} \simeq \mathcal{O}(\rho)$ for the limit $\rho \to 0$.
Finally,
inserting these expansions into Eq.(\ref{eq:fine_tune_2}),
we can find the relation with the model parameters $\alpha$ and $\mathcal{B}$:
\begin{equation}
\mathcal{B} = - \frac{1}{2\alpha} + \varXi\left[ \Tilde{f}, \Tilde{a}, \mathcal{M}, \mathcal{L} \right].
\label{eq:rel}
\end{equation}
As mentioned in the above discussion,
the relation is not useful since it is given by the functional form.
It may cause a serious problem especially on the stability analysis describing later.

\section{Numerical solutions}\label{sc:ns}

\subsection{Method}
In this subsection the numerical strategy for our system will be outlined.
An arbitrary second order dynamical system,
in principle, can be expressed in terms of a first order system {\cite{William:1995nr,Keller:1992tb,Stoer:2002na}}.
Thus we transform our equations of motion (\ref{eq:scalar})-(\ref{eq:eeq_r}) into first order differential equations.
By linearly combining the \emph{dynamical} equations (\ref{eq:scalar})-(\ref{eq:eeq_theta}),
the following set of equations can be obtained
\begin{align}
&y^{\prime}_{1}(\rho) = y_{5}(\rho),\;\;\;
y^{\prime}_{2}(\rho) = y_{6}(\rho),
\notag\\
&y^{\prime}_{3}(\rho) = y_{3}(\rho)y_{7}(\rho),\;\;\;
y^{\prime}_{4}(\rho) = y_{8}(\rho),
\notag\\
&y^{\prime}_{5}(\rho) = - \left[4y_{7}(\rho) + \frac{y_{8}(\rho)}{y_{4}(\rho)} \right]y_{5}(\rho)
\notag\\
&\phantom{y^{\prime}_{5}(\rho) =}
+ 2\left[ \frac{v_{f}(\rho)}{y^{2}_{4}(\rho)} + y_{9}(\rho)\sin{y_{1}(\rho)} \right],
\notag\\
&y^{\prime}_{6}(\rho) = - \left[ 4y_{7}(\rho) - \frac{y_{8}(\rho)}{y_{4}(\rho)} \right]y_{6}(\rho) + v_{a}(\rho),
\notag\\
&y^{\prime}_{7}(\rho) = - \frac{5}{2}y^{2}_{7}(\rho) + \frac{\alpha}{4}\left[ \tau_{\theta}(\rho) - \beta \right],
\notag\\
&y^{\prime}_{8}(\rho) = \frac{3}{2}y_{4}(\rho)y^{2}_{7}(\rho) - 3y_{7}(\rho)y_{8}(\rho)
\notag\\
&\phantom{y^{\prime}_{8}(\rho) =}
+ \frac{\alpha}{4}y_{4}(\rho)\left[ 4\tau_{0}(\rho) - 3\tau_{\theta}(\rho) - \beta \right],
\notag\\
&y^{\prime}_{9}(\rho) = 0,
\label{eqs:ne}
\end{align}
in which we introduce functions $y_{i}(\rho)\;(i = 1,\dots,9)$
defined as 
\begin{align}
&y_{1}(\rho) := \Tilde{f}(\rho),&&y_{2}(\rho) := \Tilde{a}(\rho),&&y_{3}(\rho) := \mathcal{M}(\rho),
\notag\\
&y_{4}(\rho) := \mathcal{L}(\rho),&&y_{5}(\rho) := \Tilde{f}^{\prime}(\rho),&&y_{6}(\rho) := \Tilde{a}^{\prime}(\rho),
\notag\\
&y_{7}(\rho) := m(\rho),&&y_{8}(\rho) := \mathcal{L}^{\prime}(\rho),&&y_{9}(\rho) := \gamma.
\label{eqs:nmvs}
\end{align}

Let us note that $y_{9}$ is actually not a function of $\rho$.
The reason why we temporarily regard $\gamma$ as a function of $\rho$ is as follows:
the requirement of regular geometry $\epsilon(\alpha, \beta, \gamma) = 0$ implies that $\gamma$ is a function of $\alpha$ and $\beta$,
\emph{i.e.}, $\gamma = \gamma(\alpha, \beta)$,
which means that $\gamma$ must be uniquely determined once we shall give $\alpha$ and $\beta$.
In the sense,
it is better to treat $\gamma$ as an eigenvalue of our model rather than a free parameter.
We find that $\gamma$ becomes a constant once proper numerical integrations shall be attained. 

The authors of Refs.{\cite{Peter:2003zg,Giovannini:2001hh,Roessl:2002rv}} have already pointed out that
a 3-brane model described by certain field-theoretical Lagrangian almost contains enormous numerical difficulties.
In language of a numerical analysis,
the components of Einstein equation (\ref{eq:eeq_4d})-(\ref{eq:eeq_r}) belong to the numerical class called the \emph{stiff equations}
which have exponentially growing solutions together with converging ones.
Besides solving two-point boundary-value problem is another difficulty.

The simple shooting method (SSM) is a very famous,
relatively easy but an efficient method.
Essentially SSM is a hybrid system of two solvers:
one is for solving differential equations as the initial-value problem and the other is for matching boundary conditions.
SSM is a simple and useful tool, however, necessarily must solve the initial-value problem in the interior,
some instability may occur if the solutions strongly depend on the initial conditions.
This is exactly the case that we must treat.
Solving the equations of motion (\ref{eq:scalar})-(\ref{eq:eeq_r}) by means of SSM is thus a quite difficult task.
As in the Refs.{\cite{Giovannini:2001hh,Roessl:2002rv}},
we could also achieve a localizing gravity solution using SSM combined with the Down-hill simplex method {\cite{William:1995nr}}.
At the present paper,
we employ the parallel or multiple shooting method (PSM) {\cite{Roessl:2002rv,Keller:1992tb,Stoer:2002na}}
as the solver of the two-point boundary-value problem since PSM successfully dissolves the instability of SSM very well.

The word of \emph{instability} in our problem means that
the functions released from a boundary-point tend to be divergent before arriving at the other.
In PSM, in order to evade the instability of SSM,
the integral region is divided into some subintervals where the running solutions is regular. 
After that, we apply SSM to each of them and
continue the calculation until the functions converge at all junction and both boundary-points.
Since PSM requires many initial conditions for all junction, called an initial trajectory,
if one prepares an initial trajectory approximate to proper solutions,
it significantly saves the computing time.
Thus the procedure to obtain solutions is as follows:
(i) solve the equations by SSM with the Down-hill simplex method,
(ii) use the solutions as an initial trajectory for PSM,
(iii) solve by PSM to obtain new solutions,
and (iv) repeat (ii)-(iii) until the convergence is attained.

Another problem that we should care about is that the equations are so called the over-determined series.
Our system is naturally constituted by the five equations (\ref{eq:scalar})-(\ref{eq:eeq_r}) but a number of the dynamical variables is only four.
That means Eq.(\ref{eq:eeq_r}) works as a constraint of the system.
Eq.(\ref{eq:eeq_r}) is usually used to check convergence of the numerical integration.
Also it may be a guide for finding the asymptotic solutions of the warp factors at infinity (in Sec.{\ref{sbsc:vacuum}}).
However its information is never used in Eqs.(\ref{eqs:ne}). 
If one properly takes some linear combination of the equations, 
all the five equations (\ref{eq:scalar})-(\ref{eq:eeq_r}) are involved in the numerical system. 
After that,
\begin{align}
&4m^{\prime} - \frac{20}{3}m\ell = \frac{\alpha}{3}\left( 3\tau_{\theta} - 5\tau_{r} + 2\beta \right),
\notag\\
&\ell^{\prime} + \ell^{2} + 4m\ell = \frac{\alpha}{4}\left( 4\tau_{0} + \tau_{r} - 3\tau_{\theta} - 2\beta \right)
\notag
\end{align}
or in numerical expression using Eq.(\ref{eqs:nmvs})
\begin{align}
&y_{7}^{\prime}(\rho) = \frac{5}{3}y_{7}(\rho)\frac{y_{8}(\rho)}{y_{4}(\rho)} + \frac{\alpha}{12}\left[ 3\tau_{\theta}(\rho) - 5\tau_{r}(\rho) + 2\beta \right],
\notag\\
&y_{8}^{\prime}(\rho) = - 4y_{7}(\rho)y_{8}(\rho)
\notag\\
&\phantom{y_{8}^{\prime}(\rho) =}
+ \frac{\alpha}{4}y_{4}(\rho)\left[ 4\tau_{0}(\rho) + \tau_{r}(\rho) - 3\tau_{\theta}(\rho) - 2\beta \right].
\label{eqs:new_ne}
\end{align}
are found to significantly improve the numerical convergence because the information of fifth equation is taken into account. 
Thus the system can properly realize the asymptotic behavior at infinity.
Henceforth we shall use Eqs.(\ref{eqs:new_ne}) instead of the correspondences in Eq.(\ref{eqs:ne}).

\subsection{Results}
\begin{figure}[htb]
\centering
\includegraphics[clip,scale=.9]{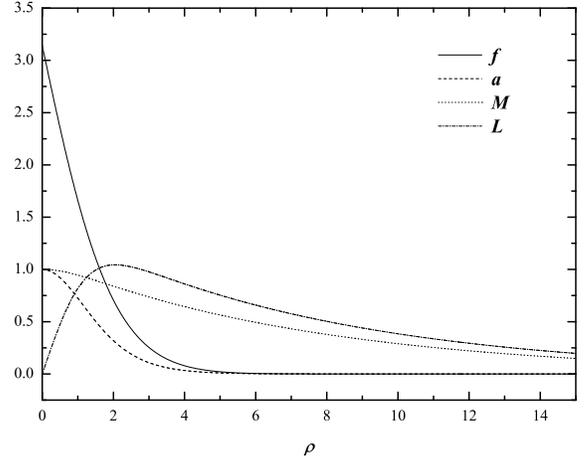}
\caption{
A typical example of the solution leading gravity localization
with the rescaled bulk radius $\rho = \sqrt{kg}r$.
The condition of shooting-parameters is as follow:
$n = 1$,
$\alpha = 0.9000000$,
$\beta = - 0.2000000$,
$\gamma = 0.488059791109$,
$\mathcal{A} = - 1.60653214371$
and $\mathcal{B} = - 0.335537911342$.
}
\label{fg:profile}
\end{figure}
\begin{figure}[htb]
\centering
\includegraphics[clip,scale=.9]{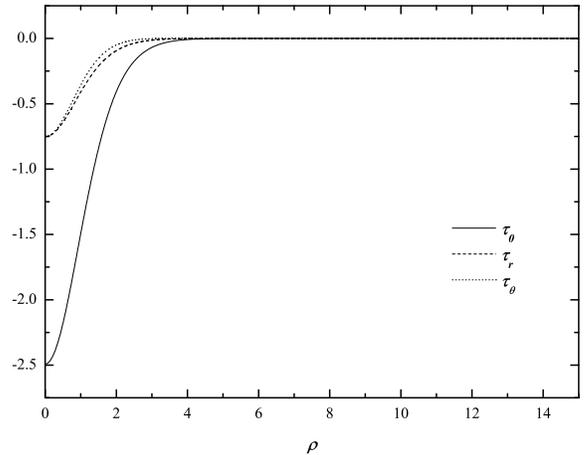}
\caption{
An example of the energy-momentum tensor for the solution given by Fig.{\ref{fg:profile}}. }
\label{fg:emt}
\end{figure}
\begin{figure}[htb]
\centering
\includegraphics[clip,scale=.9]{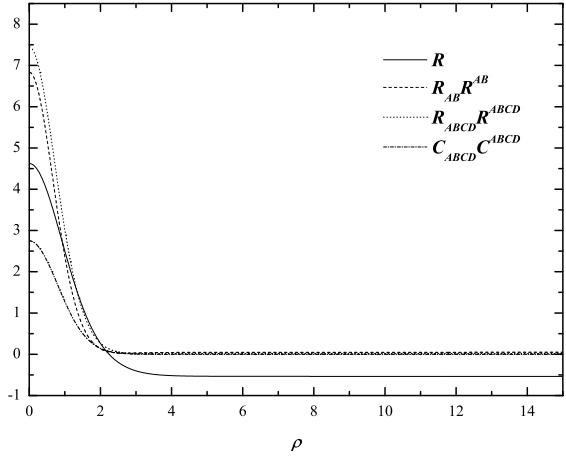}
\caption{
An example of the curvature invariants for the solution given by Fig.{\ref{fg:profile}}.
From the considerations in Sec.{\ref{sbsc:vacuum}},
we interpret that the geometry achieves AdS$_{6}$ where all curvature invariants are constant.}
\label{fg:cis}
\end{figure}
In this subsection, we present some typical examples in our numerical results.

Fig.{\ref{fg:profile}} shows a typical numerical solution realizing the gravity localization around the 3-brane.
Fig.{\ref{fg:emt}} and Fig.{\ref{fg:cis}} are the results of the energy-momentum tensor and the curvature invariants corresponding to Fig.{\ref{fg:profile}}.
The results in Fig.{\ref{fg:emt}} suggest that our solution describes a \emph{local} topological defect
because all the component of the energy-momentum tensor vanish at large $\rho$.
Furthermore,
Fig.{\ref{fg:cis}} certainly means that the geometry turns to be AdS$_{6}$ at large $\rho$.
As already mentioned in Ref.{\cite{Karch:2000ct}},
the gravity localization around a 3-brane, \emph{e.g.}, our configuration,
may be a locally trivial incident of various cosmological events in the higher dimensional world. 

\begin{figure}[htb]
\centering
\includegraphics[clip,scale=.68]{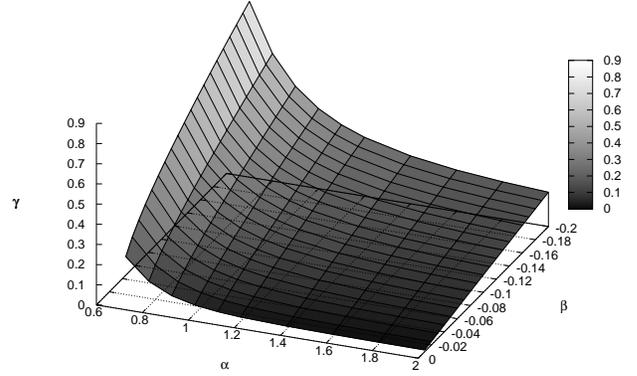}
\caption{
Parameter space for typical solutions which exactly localize gravity around the Maxwell gauged $\mathbb{C}P^{1}$ brane with the winding number $n = 1$. 
Here $\alpha=\chi k$ is the dimensionless gravitational coupling constant,
$\beta=\Lambda_{b}/k^{2}g$ is the dimensionless bulk cosmological constant
and $\gamma=\mu/k^{2}g$ is the dimensionless coupling constant of the potential like a mass term of a soliton.}
\label{fg:pspace}
\end{figure}

\begin{figure*}[t]
\centering
\includegraphics[clip,scale=1.2]{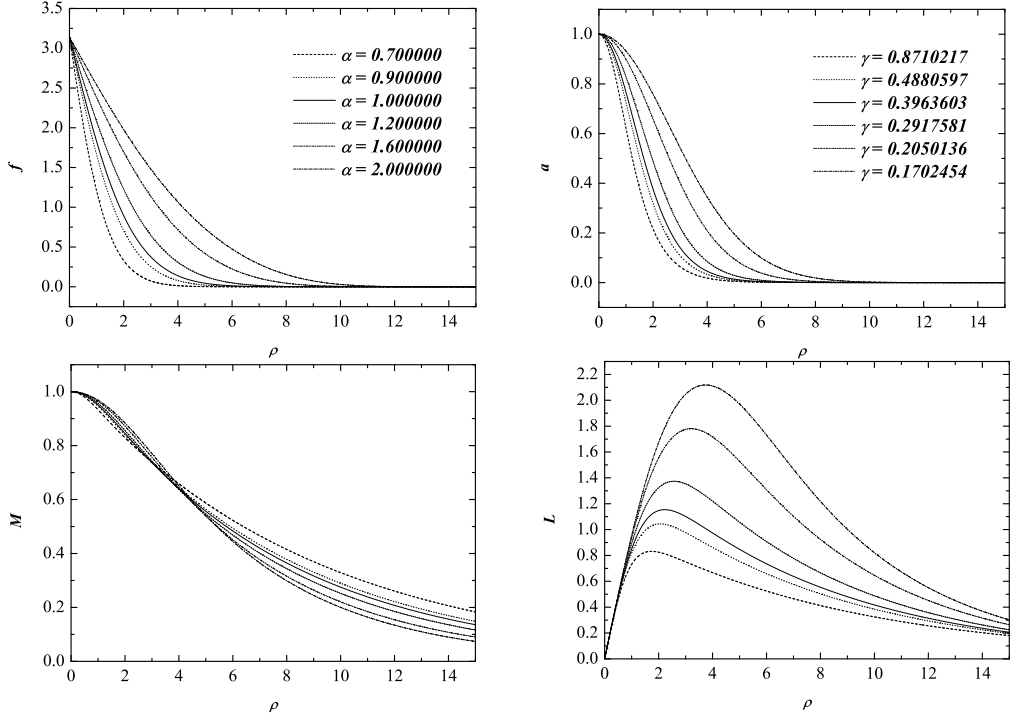}
\caption{
The behavior of the fields $\Tilde{f}(\rho), \Tilde{a}(\rho), \mathcal{M}(\rho)$ and $\mathcal{L}(\rho)$
for varying $\alpha$ at fixed $\beta = - 0.2000000$.}
\label{fg:pro_var_a}
\end{figure*}

\begin{figure*}[t]
\centering
\includegraphics[clip,scale=1.2]{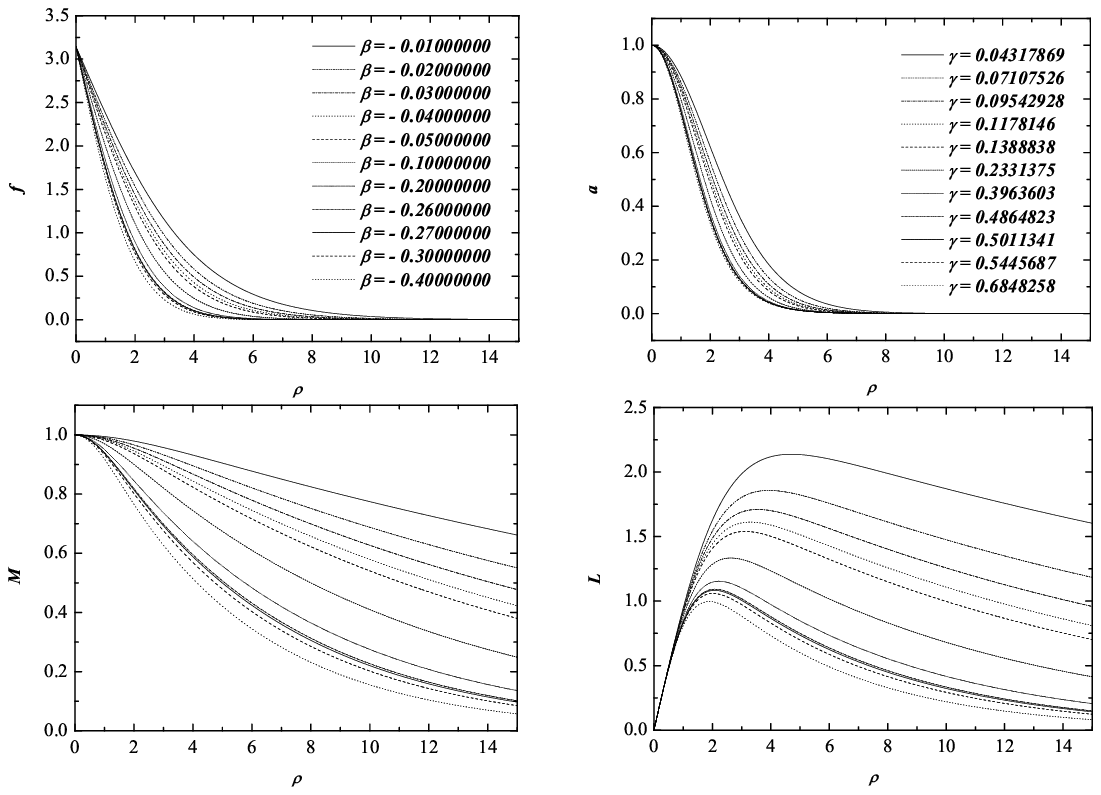}
\caption{
The behavior of the fields $\Tilde{f}(\rho), \Tilde{a}(\rho), \mathcal{M}(\rho)$ and $\mathcal{L}(\rho)$
for varying $\beta$ at fixed $\alpha = 1.000000$.}
\label{fg:pro_var_b}
\end{figure*}

Also,
Fig.{\ref{fg:pspace}} shows the fine-tuning surface in the parameter space $(\alpha, \beta, \gamma)$
corresponding to solutions for the gravity localization. 
Contrary to the seven-dimensional case {\cite{Roessl:2002rv}},
only $\beta < 0$, \emph{i.e.},
the negative bulk cosmological constant is allowed for localizing gravity solutions.
The surface is growing for $\alpha \to 0$ and $\beta \to -\infty$.
In numerous studies {\cite{Giovannini:2001hh,Roessl:2002rv,Peter:2003zg}},
the authors have insisted that the property of the solutions are dominated by the parameter $c$, \emph{i.e.}, Eq.(\ref{eq:parameter_c})
which relates to the asymptotic behavior of the warp factors at infinity and of course has much importance.
Fig.{\ref{fg:pspace}} tells us, however,
that the dependence of the surface on the parameters $\alpha$ and $\beta$ is complicated,
thus more thorough analysis for physical implication of the parameter-dependence is required (see also Fig.{\ref{fg:pro_var_a}} and Fig.{\ref{fg:pro_var_b}}).
In fact,
$\gamma = \gamma(\alpha,\beta)$ is the dimensionless coefficient of the potential term and
the variation of $\gamma$ brings remarkable change of the property of the solutions.

\section{Linear stability analysis}\label{sc:lgp}
In order to clarify the physical implication of our brane solutions, 
we proceed to linear stability analysis concerning the solutions. 
The stability of the solutions remains an open problem though it may be generally guaranteed by its topology.
In most of the previous works {\cite{Peter:2003zg,Giovannini:2002sb,Giovannini:2002mk,Giovannini:2002jf,Giovannini:2001xg}},
the source of gravity is well known Abelian vortex which is described by a scalar singlet and $U(1)$ gauge field.
The stability for the models in those works is quasi-stable,
which means that brane models described by topological solitons are not always engaged its stability.
On the other hand,
$\mathbb{C}P^{1}$ model is written in terms of a scalar doublet (or a scalar triplet in terms of $O(3)$ sigma model) and is not based on gauge theory.
Study of linear stability analysis for the present model is absent.
It is thus worthwhile to examine its linear stability afresh.

\subsection{Gauge-invariant framework}
The perturbed linearized Einstein equation is obtained by performing the first-order perturbation of the background metric tensor.
This procedure is straightforward,
but one should care about the freedom of gauge,
\emph{i.e.}, the choice of background coordinates before beginning the analysis.
We adopt a gauge-invariant approach called the \emph{longitudinal} gauge choice.
The detailed discussion is given by Appendix {\ref{ap:gifl}}.
In this subsection we shall show only the outline of the analysis.

The gauge-invariant fluctuations for the original metric tensor (\ref{eq:metric}) can be written as
\begin{equation}
\delta\mathcal{G}_{MN} = \begin{pmatrix}
2M^{2}\mathcal{U}_{\mu\nu} & M\mathcal{E}_{\nu} & ML\mathcal{F}_{\nu}  \cr
M\mathcal{E}_{\mu} & 2\mathcal{X} & L\varOmega \cr
ML\mathcal{F}_{\mu} & L\varOmega & 2L^{2}\varPhi \cr
\end{pmatrix},
\label{eq:gi_pmet}
\end{equation}
where
\begin{equation}
\mathcal{U}_{\mu\nu} = h_{\mu\nu} + \eta_{\mu\nu}\varPsi
\label{eq:gi_pmet_4d}
\end{equation}
is the four-dimensional components.
Also the gauge-invariant perturbations of the matter fields (\ref{eqs:matter_ansatz}) is defined by
\begin{align}
&\delta\bm{z} = \left( \varSigma_{1}e^{-in\theta},\; \varSigma_{2} \right)^{T},
\notag\\
&\delta A_{M} = \left( \varTheta_{\mu} + \partial_{\mu}\varTheta,\; \varTheta_{r},\; \varTheta_{\theta} \right).
\label{eqs:gi_pmatter}
\end{align}
In the above equations,
all the perturbed fields are functions of the \emph{full} spatial coordinates $x^{M}=(x^{\mu},r,\theta)$.

In order to derive the evolution equations of the fluctuations (\ref{eq:gi_pmet})-(\ref{eqs:gi_pmatter}), 
it is convenient to introduce some perturbed quantities,
\emph{e.g.}, the first order perturbation of Einstein tensor
\begin{equation}
\delta G_{AB} = \delta R_{AB} - \frac{1}{2}
\left( \mathcal{G}_{AB}\delta R + \delta \mathcal{G}_{AB}R \right),
\label{eq:p_etensor}
\end{equation}
where
\begin{equation}
\delta R = \mathcal{G}^{AB}\delta R_{AB} + \delta \mathcal{G}^{AB}R_{AB}
\label{eq:p_rscalar}
\end{equation}
is the perturbed Ricci scalar curvature.
Furthermore,
if one would like to find the explicit form of Eqs.(\ref{eq:p_etensor})-(\ref{eq:p_rscalar}), 
one must calculate other perturbed quantities
\begin{equation}
\delta\mathcal{G}^{MN} = -\mathcal{G}^{MA}\mathcal{G}^{NB}\delta\mathcal{G}_{AB}
\label{eq:p_imet}
\end{equation}
for the inverse metric tensor,
\begin{equation}
\delta \Gamma_{AB}^{C} = \frac{1}{2}\mathcal{G}^{CD}
\left( \nabla_{A}\delta\mathcal{G}_{BD} 
+ \nabla_{B}\delta\mathcal{G}_{DA} - \nabla_{D}\delta\mathcal{G}_{AB} \right)
\label{eq:p_christ}
\end{equation}
for the Christoffel connection and
\begin{multline}
\delta R^{A}_{BCD} =
- \partial_{D}\delta\Gamma^{A}_{BC}
+ \partial_{C}\delta\Gamma^{A}_{BD}
+ \delta\Gamma^{E}_{BD}\Gamma^{A}_{EC}\\
+ \Gamma^{E}_{BD}\delta\Gamma^{A}_{EC}
- \delta\Gamma^{E}_{BC}\Gamma^{A}_{ED}
- \Gamma^{E}_{BC}\delta\Gamma^{A}_{ED}
\label{eq:p_riemann}
\end{multline}
for the Riemann tensor. 
In Eq.(\ref{eq:p_christ}),
$\nabla_{M}$ denotes the covariant derivative with respect to the metric tensor (\ref{eq:metric}).
The evolution equations given by fluctuating the original equations of motion (\ref{eqs:eqs}) are thus obtained
\begin{align}
&\delta G_{MN} = \chi\left(\delta T_{MN} - \Lambda_{b}\delta\mathcal{G}_{MN}\right),
\notag\\
&\delta\nabla_{M}\left(\mathcal{D}^{M}z^{a}\right)+ i\delta\left(A_{M}\mathcal{D}^{M}z^{a}\right) 
+ \frac{\mu}{k}\sigma_{3}^{ab}\delta z^{b} = 0,
\notag\\
&\delta\nabla_{M}F^{MN} = 8kg\delta\left(A^{N}\left|\bm{z}\right|^{2}\right)
\notag\\
&\phantom{\delta\nabla_{M}F^{MN} =}
+ i4kg\delta\left(z^{a}\partial^{N}z^{a*} - z^{a*}\partial^{N}z^{a}\right),
\label{eqs:per_eqs}
\end{align}
where
\begin{multline}
\delta T_{MN} =
2k \delta \left[ \left( \mathcal{D}_{(M}z^{a} \right)^{*} \mathcal{D}_{N)}z^{a} \right]
+ \frac{1}{g} \delta \left[ \mathcal{G}^{AB} F_{MA}F_{MB} \right]\\
+ \left( \delta \mathcal{G}_{MN} \right) \mathcal{L}_{\mathrm{brane}}
+ \mathcal{G}_{MN} \delta \mathcal{L}_{\mathrm{brane}},
\label{eq:gp_emt}
\end{multline}
is the perturbed energy-momentum tensor with the perturbed brane Lagrangian density
\begin{multline}
\delta\mathcal{L}_{\mathrm{brane}} =
- k \delta\left[\left( \mathcal{D}_{M}z^{a} \right)^{*} \mathcal{D}^{M}z^{a}\right]
- \frac{1}{4g} \delta\left(F_{MN}F^{MN}\right)\\
- \mu \left[ 1 - \delta\left(\bm{z}^{\dag} \bm{\sigma}_{3} \bm{z}\right) \right].
\label{eq:pb_lagrangian}
\end{multline}
$X_{(M}Y_{N)}$,
in the first term of the right-hand-side of Eq.(\ref{eq:gp_emt}),
means as the symmetrization {\cite{Carmeli:2001ay}} defined by
\begin{equation}
X_{(M}Y_{N)} := \frac{1}{2!}\left( X_{M}Y_{N} + X_{N}Y_{M} \right).
\notag
\end{equation}
Since the evaluation of the evolution equations (\ref{eqs:per_eqs})-(\ref{eq:pb_lagrangian}) is straightforward but includes tedious algebra, 
we shall show the detailed descriptions in Appendix {\ref{ap:pfes}}.

Note that the symmetrization in Eq.(\ref{eq:gp_emt}) naturally realizes the real gauge condition {\cite{Peter:2003zg}}. 
The $\varSigma_{i}$ ($i = 1,2$) may be, in general, complex quantities, 
but we can exclude the imaginary part of $\varSigma_{i}$ by applying certain $U(1)$ gauge rotation.
In mathematical point of view,
the symmetrization is equivalent to choosing the real part of a complex values.

\subsection{Tensor zero mode fluctuation}
For the tensor mode fluctuation,
only the $(\mu,\nu)$ component of the perturbed Einstein equation is sufficient for the analysis.
After a lengthy algebra,
the final result of a evolution equation for the tensor mode fluctuation $h_{\mu\nu}$ is given by
\begin{equation}
\frac{1}{M^{2}}\square h_{\mu\nu}
+ h_{\mu\nu}^{\prime\prime}
+ \frac{1}{L^{2}}\Ddot{h}_{\mu\nu}
+ \left( 4\Hat{m} + \Hat{\ell} \right)h_{\mu\nu}^{\prime} = 0,
\label{eq:peq_h}
\end{equation}
where $\square$ is the flat four-dimensional d'Alembertian
\begin{equation}
\square = \eta^{\mu\nu}\partial_{\mu}\partial_{\nu} = - \partial_{t}^{2} + \sum_{i=1}^{3}\partial_{i}^{2}
\label{eq:4d_d'Alembertian}
\end{equation}
and
\begin{equation}
\Hat{m} := \frac{d\ln{M(r)}}{dr},\;\;\; \Hat{\ell} := \frac{d\ln{L(r)}}{dr}.
\label{eqs:CH_trans}
\end{equation}
In Eq.(\ref{eq:peq_h}) the prime and the over-dot denote the derivative with respect to the bulk radius $r$ and the bulk angle $\theta$, respectively.
The first term in Eq.(\ref{eq:peq_h}) can be $\square h_{\mu\nu} = m_{h}^{2}h_{\mu\nu}$ for the massive graviton,
since $h_{\mu\nu}$ satisfies the four-dimensional Klein-Gordon equation.
The term vanishes for the massless graviton.

Any physical quantity must exhibit axial symmetry since the geometry has same symmetry.
We restrict $h_{\mu\nu}$ to the massless ($m_{h}=0$) and the zero modes ($p=0$), 
then
\begin{equation}
h_{\mu\nu}(x^{M}) = \sum_{p\in\mathbb{Z}}h_{\mu\nu}(x^{\lambda},r)e^{ip\theta} \to h_{\mu\nu}(r).
\notag
\end{equation}
Eq.(\ref{eq:peq_h}) is fairly simplified to 
\begin{equation}
\Tilde{h}_{\mu\nu}^{\prime\prime}(\rho) + \left( 4m + \ell \right)\Tilde{h}_{\mu\nu}^{\prime}(\rho) = 0,
\notag
\end{equation}
where
\begin{equation}
\Tilde{h}_{\mu\nu}(\rho) := h_{\mu\nu}(r)
\notag
\end{equation}
is the dimensionless quantity
and the prime, \emph{acting on variables with the tilde}, denotes the derivative with respect to $\rho$.
Also the functions $m$ and $\ell$ in the equation are defined by Eq.(\ref{eqs:dimensionless_CH_trans}).
One easily sees that an arbitrary constant is allowed as the solution of $\Tilde{h}_{\mu\nu}$.
Thus we employ $\Tilde{h}_{\mu\nu}(\rho) := h$ (constant).
Since a normalizability of $\Tilde{h}_{\mu\nu}(\rho)$ is essentially equivalent to the integral (\ref{eq:planckmass}),
it is automatically guaranteed if the background fields of perturbation satisfy 
the finiteness condition of the four-dimensional Planck mass (\ref{eq:planckmass}).

Some authors in this field have arrived at similar conclusions {\cite{Giovannini:2002sb,Giovannini:2002mk}}.
Since the tensor mode $\Tilde{h}_{\mu\nu}$ does not couple with the source term of gravity, 
properties of the localized massless graviton around a topological defect may be independent on the gravitational source in models.

\subsection{Vector zero mode fluctuations}
Evolution equations for the vector mode fluctuations 
$\mathcal{E}_{\mu}, \mathcal{F}_{\mu}$ and $\varTheta_{\mu}$ are derived
from the $(\mu, \nu)$, $(\mu,r)$ and $(\mu,\theta)$ components of the Einstein equation
and also the $(\mu)$ component of the equation of the $U(1)$ gauge field. 
Here we introduce two new variables for convenience
\begin{equation}
\mathfrak{C}_{\mu} := \frac{\mathfrak{K}}{L}\Dot{\mathcal{E}}_{\mu} 
- \left( \mathfrak{K}\mathcal{F}_{\mu} \right)^{\prime},\;\;\;
\mathfrak{K} := \frac{L}{M}.
\label{eqs:sa_v_nv}
\end{equation}
Similar replacements already have been used in Refs.{\cite{Giovannini:2002sb,Giovannini:2002mk}}.
From these variables,
we obtain four equations
\begin{align}
&\mathcal{E}_{\mu}^{\prime} + \left( 4\Hat{m} 
+ \frac{\mathfrak{K}^{\prime}}{\mathfrak{K}} \right)\mathcal{E}_{\mu} + \frac{1}{L}\Dot{\mathcal{F}}_{\mu} = 0,
\label{eq:peq_v_constraint}\\
&\frac{1}{M^{2}}\square\mathcal{E}_{\mu} + \frac{1}{\mathfrak{K}L}\Dot{\mathfrak{C}}_{\mu}
+ \frac{\chi}{g}\frac{2a^{\prime}}{ML^{2}}\Dot{\varTheta}_{\mu} = 0,
\label{eq:peq_v_e}\\
&\frac{1}{M^{2}}\square\mathcal{F}_{\mu} - \frac{1}{\mathfrak{K}}\mathfrak{C}_{\mu}^{\prime}
- \left( 5\Hat{m} - \frac{\mathfrak{K}^{\prime}}{\mathfrak{K}} \right)\frac{1}{\mathfrak{K}}\mathfrak{C}_{\mu}
\notag\\
&\phantom{\frac{1}{M^{2}}\square\mathcal{F}_{\mu}}
- \frac{2\chi}{ML}\left( \frac{a^{\prime}}{g}\varTheta_{\mu}^{\prime} + kv_{a}\varTheta_{\mu} \right) = 0,
\label{eq:peq_v_f}\\
&\frac{1}{M^{2}}\square\varTheta_{\mu} + \varTheta_{\mu}^{\prime\prime} + \frac{1}{L^{2}}\Ddot{\varTheta}_{\mu}
\notag\\
&\phantom{\frac{1}{M^{2}}\square\varTheta_{\mu}}
+ \left( 3\Hat{m} + \frac{\mathfrak{K}^{\prime}}{\mathfrak{K}} \right)\varTheta_{\mu}^{\prime}
- \frac{a^{\prime}}{\mathfrak{K}^{2}}\mathfrak{C}_{\mu} = 8kg\varTheta_{\mu}.
\label{eq:peq_v_t}
\end{align} 

We again concentrate on the lowest angular momentum eigenstate upon the three vector modes.
Then Eq.(\ref{eq:peq_v_e}) turns to be
\begin{equation}
\frac{1}{M^{2}}\square\mathcal{E}_{\mu}
= \left. \frac{1}{M^{2}}\sum_{p\in\mathbb{Z}}\square\mathcal{E}_{\mu}(x^{\nu}, r)e^{ip\theta} \right|_{p=0} = 0
\notag
\end{equation}
which means that $\square\mathcal{E}_{\mu}(x^{\nu},r) = 0$, therefore, for all integer $p$,
\begin{equation}
\sum_{p\in\mathbb{Z}}\square\mathcal{E}_{\mu}(x^{\nu}, r)e^{ip\theta} = 0.
\notag
\end{equation}
Then we conclude the vector mode $\mathcal{E}_{\mu}$ should be always the massless graviphoton.
Inserting the condition into Eq.(\ref{eq:peq_v_e}),
we obtain the important relation
\begin{equation}
\mathfrak{C}_{\mu} = - \frac{2\chi}{g}\frac{a^{\prime}}{M^{2}}\varTheta_{\mu}
\label{eq:rel_c_t}
\end{equation}
which is valid for \emph{general} angular momentum eigenstates
since the vector mode $\mathcal{E}_{\mu}$ is the massless graviphoton.
Also using the relation into Eq.(\ref{eq:peq_v_f}),
we obtain
\begin{align}
\frac{1}{M^{2}}\square\mathcal{F}_{\mu} &=
- \frac{2\chi}{g}\frac{1}{ML}\left[ a^{\prime\prime} 
+ \left( 4\Hat{m} - \Hat{\ell} \right)a^{\prime} - kgv_{a} \right]\varTheta_{\mu}
\notag\\
&= 0.
\notag
\end{align}
For the last line we used the equation for the $U(1)$ gauge field (\ref{eq:gauge}).
Consequently, the vector mode $\mathcal{F}_{\mu}$ should be also the massless graviphoton for arbitrary angular momentum eigenstates.
Those results mean, from Eqs.(\ref{eqs:sa_v_nv}) and (\ref{eq:rel_c_t}),
the vector mode $\varTheta_{\mu}$ should be always the massless gauge degree of freedom too.

Next we insert Eq.(\ref{eq:rel_c_t}) into Eq.(\ref{eq:peq_v_t}) 
and define a dimensionless function with the rescaled variable $\rho$
\begin{equation}
\Tilde{\varTheta}_{\mu}(\rho) := \frac{\varTheta_{\mu}(r)}{\sqrt{kg}},
\notag
\end{equation}
we obtain the evolution equation for the rescaled vector mode $\Tilde{\varTheta}_{\mu}(\rho)$
\begin{multline}
\Tilde{\varTheta}_{\mu}^{\prime\prime}(\rho) + \left( 2m + \ell \right)\Tilde{\varTheta}_{\mu}^{\prime}(\rho)\\
+ 2\left( \alpha\frac{\Tilde{a}^{\prime2}}{\mathcal{L}} - 4 \right)\Tilde{\varTheta}_{\mu}(\rho) = 0.
\label{eq:peq_v_t_rescaled}
\end{multline}
In the analysis of Abelian vortex {\cite{Giovannini:2002sb,Giovannini:2002mk}},
all evolution equations for the vector mode fluctuation have been solvable
because the useful relation for the string tension was found.
Unfortunately,
in our case Eq.(\ref{eq:peq_v_t_rescaled}) is not analytically solvable
since the similar relation (\ref{eq:fine_tune}) are less tractable for the analysis,
as stated in Sec.{\ref{sc:as}}. 
In this paper,
we shall examine asymptotic solutions of $\Tilde{\varTheta}$ at both boundaries.
At infinity,
since we can linearize the evolution equation (\ref{eq:peq_v_t_rescaled}),
we are able to find the asymptotic equation
\begin{equation}
\Tilde{\varTheta}_{\mu}^{\prime\prime}(\rho) - 3c\,\Tilde{\varTheta}_{\mu}^{\prime}(\rho) - 8\Tilde{\varTheta}_{\mu}(\rho) = 0
\notag
\end{equation}
and the solution
\begin{equation}
\Tilde{\varTheta}_{\mu}(\rho)  \propto \exp\left[ - \frac{3c}{2}\left( -1 \pm \sqrt{1 + \frac{32}{9c^{2}}} \right)\rho \right].
\label{eq:asymp_t_inf}
\end{equation}
Also at the vicinity of the origin,
using the asymptotic solutions (\ref{eqs:asymp_origin}) for the background of perturbation
and imposing a regularity of Eq.(\ref{eq:peq_v_t_rescaled}) at the origin,
we find the asymptotic solution at the other boundary
\begin{equation}
\Tilde{\varTheta}_{\mu}(\rho) \simeq \vartheta_{\mu}^{(0)}\left[ 1 + 2\left( 1 - \alpha\mathcal{B}^{2} \right)\rho^{2} \right] + \mathcal{O}\left( \rho^{3} \right),
\label{eq:asymp_t_origin}
\end{equation}
where $\vartheta_{\mu}^{(0)}$ is an arbitrary constant.
From them,
we speculate that $\Tilde{\varTheta}_{\mu}$ is almost localizing around the defect.

For the vector mode $\mathcal{E}_{\mu}$,
we easily obtain the analytical solution from the constraint (\ref{eq:peq_v_constraint}).
After rescaling the two vector modes with respect to $\rho$, \emph{i.e.},
\begin{equation}
\Tilde{\mathcal{E}}_{\mu}(\rho) := \mathcal{E}_{\mu}(r),\;\;\;
\Tilde{\mathcal{F}}_{\mu}(\rho) := \mathcal{F}_{\mu}(r),
\notag
\end{equation}
we immediately achieve the equation
\begin{equation}
\Tilde{\mathcal{E}}_{\mu}^{\prime}(\rho) + \left( 3m + \ell \right)\Tilde{\mathcal{E}}_{\mu}(\rho) = 0
\notag
\end{equation}
with the solution
\begin{equation}
\Tilde{\mathcal{E}}_{\mu}(\rho) = \frac{e_{\mu}^{(0)}}{\mathcal{M}^{3}\mathcal{L}},
\notag
\end{equation}
where $e_{\mu}^{(0)}$ is an integration constant.
For the vector mode $\mathcal{F}_{\mu}$,
from Eq.(\ref{eq:rel_c_t}),
we obtain 
\begin{equation}
\Tilde{\mathcal{F}}_{\mu}(\rho) =
\frac{\mathcal{M}}{\mathcal{L}}\left( f_{\mu}^{(0)} + 2\alpha\int{d\rho}
\frac{\Tilde{a}^{\prime}\Tilde{\varTheta}_{\mu}}{\mathcal{M}^{2}} \right),
\notag
\end{equation}
where $f_{\mu}^{(0)}$ is also an integration constant.
This solution, unfortunately, is a functional of $\Tilde{a}^{\prime}, \Tilde{\varTheta}_{\mu}$, and $\mathcal{M}$.

Finally we examine normalizability of the three vector mode fluctuations.
For $\Tilde{\mathcal{E}}_{\mu}, \Tilde{\mathcal{F}}_{\mu}$, the integrals become
\begin{equation}
\int_{0}^{\infty}d\rho\mathcal{M}^{4}\mathcal{L}\left|\Tilde{\mathcal{E}}_{\mu}(\rho)\right|^{2}
= \int_{0}^{\infty}d\rho\frac{e_{\mu}^{(0)2}}{\mathcal{M}^{2}\mathcal{L}},
\label{eq:nm_e}
\end{equation}
and
\begin{multline}
\int_{0}^{\infty}d\rho\mathcal{M}^{4}\mathcal{L}\left|\Tilde{\mathcal{F}}_{\mu}(\rho)\right|^{2}\\
= \int_{0}^{\infty}d\rho\frac{\mathcal{M}^{6}}{\mathcal{L}}
\left( f_{\mu}^{(0)} + 2\alpha\int{d\rho}\frac{\Tilde{a}^{\prime}\Tilde{\varTheta}_{\mu}}{\mathcal{M}^{2}} \right)^{2}.
\label{eq:nm_f}
\end{multline}
Clearly,
they are not normalizable if the background solutions lead localizing gravity around the 3-brane.
Moreover,
the integrand of Eq.(\ref{eq:nm_e}) is certainly divergent at the both boundaries
and that of Eq.(\ref{eq:nm_f}) is also divergent at the origin.
Thus we confirm that the two modes are not renormalizable without any information about
a explicit form of the vector mode $\Tilde{\varTheta}_{\mu}$.
On the other hand,
though we could not assure that the normalization integral for $\Tilde{\varTheta}_{\mu}$ is exactly finite, 
we confirm the integrand is finite at both boundaries from Eqs.(\ref{eq:asymp_t_inf})-(\ref{eq:asymp_t_origin}). 
In that sense, $\Tilde{\varTheta}_{\mu}$ is the potentially normalizable vector mode fluctuation.

\subsection{Scalar zero mode fluctuations}
Since all perturbed equations contain the scalar mode fluctuations,
we must deal with the whole evolution equations (\ref{eqs:per_eqs}).
(Their explicit forms are given in Appendix {\ref{ap:pfes}}.)
In Ref.{\cite{RandjbarDaemi:2002pq}},
in order to treat such complicated coupled system,
the authors divide the whole system into many distinct subsets by using some redefinitions of the fields.
As Ref.{\cite{Peter:2003zg}}, however, we only have two subsets from Eqs.(\ref{eqs:per_eqs})
if we restrict our analysis to the only lowest angular momentum eigenstates.
One of subsets, called \emph{sector I}, contains the scalar mode fluctuations $(\varTheta, \varTheta_{r}, \varOmega)$,
and another subset, called \emph{sector II}, contains $(\varPsi, \mathcal{X}, \varPhi, \varSigma_{1}, \varSigma_{2}, \varTheta_{\theta})$.

\subsubsection{Sector I: $\varTheta, \varTheta_{r}, \varOmega$}
This subset is constructed by the six evolution equations of Eqs.(\ref{eqs:per_eqs});
the $(\mu,\theta)$ and $(r,\theta)$ components of the perturbed Einstein equation are
\begin{equation}
\varOmega^{\prime} + 2\left( \Hat{m} + \Hat{\ell} \right)\varOmega
= \frac{2\chi}{g}\frac{a^{\prime}}{L}\left( \varTheta_{r} - \varTheta^{\prime} \right)
- 2k\chi\frac{v_{a}}{L}\varTheta
\label{eq:p_ein_mt}
\end{equation}
and
\begin{multline}
- \frac{1}{2M^{2}}\square\varOmega
+ \left( \Hat{m}^{\prime} - \Hat{\ell}^{\prime} + 4\Hat{m}^{2} - \Hat{\ell}^{2} - 3\Hat{m}\Hat{\ell} \right)\varOmega
\\
= \frac{\chi}{g}\frac{a^{\prime2}}{L^{2}}\varOmega - k\chi\frac{v_{a}}{L}\varTheta_{r},
\label{eq:p_ein_rt}
\end{multline}
the $(\mu)$ and $(r)$ components of the perturbed $U(1)$ gauge equation are 
\begin{align}
&\varTheta^{\prime\prime} - \varTheta_{r}^{\prime}
+ \left( 2\Hat{m} + \Hat{\ell} \right)\left( \varTheta^{\prime} - \varTheta_{r} \right)
= 8kg\varTheta,
\notag\\
&\frac{1}{M^{2}}\square\left( \varTheta_{r} - \varTheta^{\prime} \right) - \frac{\varOmega}{L}v_{a} = 8kg\varTheta_{r},
\label{eqs:p_u1_s1}
\end{align}
and two imaginary parts of the perturbed $\mathbb{C}P^{1}$ equation in Appendix {\ref{ap:pfes}}.
Since the present sector contains two constraint equations,
we can perform to stability analysis of the sector with only three equations (\ref{eq:p_ein_mt})-(\ref{eqs:p_u1_s1})

Here we define a three dimensionless variables
\begin{equation}
\Tilde{\varTheta}(\rho) := \varTheta(r),\;\;\;
\Tilde{\varTheta}_{r}(\rho) := \frac{\varTheta_{r}(r)}{\sqrt{kg}},\;\;\;
\Tilde{\varOmega}(\rho) := \varOmega(r),
\notag\\
\end{equation}
and also two new variables
\begin{equation}
\Tilde{\varTheta}_{1}(\rho) := \Tilde{\varTheta}^{\prime}(\rho) - \Tilde{\varTheta}_{r}(\rho),\;\;\;
\Tilde{\varTheta}_{2}(\rho) := \Tilde{\varTheta}(\rho).
\notag
\end{equation}
Using these variables,
we get the two equations from Eqs.(\ref{eq:p_ein_mt}) and (\ref{eqs:p_u1_s1}),
\begin{align}
&\Tilde{\varOmega}^{\prime}(\rho) + 2\left( m + \ell \right)\Tilde{\varOmega}(\rho)
= - \frac{2\alpha}{\mathcal{L}}\left( \Tilde{a}^{\prime}\Tilde{\varTheta}_{1}(\rho) + v_{a}\Tilde{\varTheta}_{2}(\rho) \right),
\notag\\
&\Tilde{\varTheta}_{1}^{\prime}(\rho) + \left( 2m + \ell \right)\Tilde{\varTheta}_{1}(\rho) = 8\Tilde{\varTheta}_{2}(\rho),
\label{eqs:pv_dynm}
\end{align}
and the two constraints from Eqs.(\ref{eq:p_ein_rt}) and (\ref{eqs:p_u1_s1}),
\begin{align}
&2v\Tilde{\varOmega}(\rho) = v_{a}\left( \Tilde{\varTheta}_{1}(\rho) - \Tilde{\varTheta}_{2}^{\prime}(\rho) \right),
\notag\\
&\frac{v_{a}}{\mathcal{L}}\Tilde{\varOmega}(\rho) = 8\left( \Tilde{\varTheta}_{1}(\rho) - \Tilde{\varTheta}_{2}^{\prime}(\rho) \right),
\label{eqs:pv_cnst}
\end{align}
in which we assumed all the three degrees of freedom are massless
and used the relation
\begin{align}
&m^{\prime} - \ell^{\prime} + 4m^{2} -\ell^{2} -3m\ell,
\notag\\
&= \alpha\left( \tau_{\theta} - \tau_{0} \right)
= \alpha\left( \frac{\Tilde{a}^{\prime2}}{\mathcal{L}^{2}} + \frac{2v}{\mathcal{L}^{2}} \right)
\notag
\end{align}
which is obtained from the components of original Einstein equation (\ref{eq:eeq_4d}) and (\ref{eq:eeq_theta}) and also the relation (\ref{eq:left_1}). 
\begin{figure}[t]
\centering
\includegraphics[clip,scale=.85]{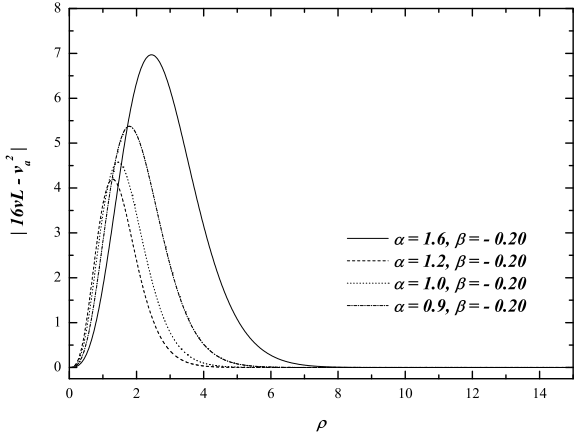}
\caption{
The magnitude of the function $| 16v\mathcal{L} - v_{a}^{2}|$ for various parameter choices.}
\label{fg:condition}
\end{figure}
In order to satisfy Eqs.(\ref{eqs:pv_cnst}),
there are two possibilities:
\begin{equation}
\frac{2v}{v_{a}} = \frac{v_{a}}{8\mathcal{L}}\;\;\;\text{or}\;\;\; \Tilde{\varOmega}(\rho) = 0.
\notag
\end{equation}
The former condition obviously can not be achieved by the solutions of the numerical integration,
\emph{e.g.}, shown in Fig.{\ref{fg:condition}},
therefore we choose $\Tilde{\varOmega}(\rho) = 0$.

From Eqs.(\ref{eqs:pv_cnst}) we find that the useful relation
\begin{equation}
\Tilde{\varTheta}_{1}(\rho) = \Tilde{\varTheta}_{2}^{\prime}(\rho).
\label{eq:rel_s1}
\end{equation} 
Inserting Eq.(\ref{eq:rel_s1}) into Eq.(\ref{eqs:pv_dynm}),
we finally attain the evolution equation
\begin{equation}
\Tilde{\varTheta}_{2}^{\prime\prime}(\rho) + \left( 2m + \ell \right)\Tilde{\varTheta}_{2}^{\prime}(\rho) - 8\Tilde{\varTheta}_{2}(\rho) = 0.
\label{eq:peq_s_t_rescaled}
\end{equation}
Similarly to the above two subsections,
this equation is not analytically solvable since $m$ and $\ell$ are the functions of $\rho$,
but we can observe the asymptotic behaviors at both boundaries.
At infinity the asymptotic solution for $\Tilde{\varTheta_{2}}$ is
\begin{equation}
\Tilde{\varTheta}_{2}(\rho) \propto \exp\left[ - \frac{3c}{2}\left( -1 \pm \sqrt{1 + \frac{32}{9c^{2}}} \right)\rho \right].
\label{eq:asym_t2_inf}
\end{equation}
From Eqs.(\ref{eq:rel_s1}) and (\ref{eq:asym_t2_inf}),
the asymptotic solution for $\Tilde{\varTheta_{1}}$ is also
\begin{equation}
\Tilde{\varTheta}_{1}(\rho) \propto \Tilde{\varTheta}_{2}(\rho).
\label{eq:asym_t1_inf}
\end{equation}
The asymptotic solutions for $\Tilde{\varTheta}_{1}$ and $\Tilde{\varTheta}_{2}$ 
are essentially the same as Eq.(\ref{eq:asymp_t_inf}).
On the other hand,
we find the asymptotic solution at the origin by expanding the equations and the functions 
as the same method in Sec.{\ref{sc:as}}.
The results are
\begin{align}
&\Tilde{\varTheta}_{1}(\rho) \simeq 4\vartheta_{2}^{(0)}\rho + \mathcal{O}(\rho^{2}),
\notag\\
&\Tilde{\varTheta}_{2}(\rho) \simeq \vartheta_{2}^{(0)}\left( 1 + 2\rho^{2} \right) + \mathcal{O}(\rho^{3})
\end{align}
where $\vartheta_{2}^{(0)}$ is an arbitrary constant.
Therefore we suggest that $\Tilde{\varTheta}_{1}$ and $\Tilde{\varTheta}_{2}$ are potentially normalizable
since the normalizability integrals for them are regular at the both boundaries. 

\subsubsection{Sector II: $\varPsi, \mathcal{X}, \varPhi, \varSigma_{1}, \varSigma_{2}, \varTheta_{\theta}$}
Analysis of the sector II is much more complicated than the sector I,
since in the sector II we should treat the six degrees of freedom and the eight evolution equations which we shall see in the below.
Note that we again restrict the analysis for the lowest angular momentum eigenstates upon all the quantities.
Firstly for the five components of the perturbed Einstein equation belonging to the present sector:
the $(\mu \ne \nu)$ component is
\begin{equation}
\partial_{\mu}\partial_{\nu}\left( 2\varPsi + \mathcal{X} + \varPhi \right) = 0,
\label{eq:p_ein_mn_s2}
\end{equation}
the $(\mu = \nu)$ component is
\begin{widetext}
\begin{multline}
3\varPsi^{\prime\prime} + \varPhi^{\prime\prime}
+ 3\left( 4\Hat{m} + \Hat{\ell} \right)\varPsi^{\prime}
- \left( 3\Hat{m} + \Hat{\ell} \right)\left( \mathcal{X}^{\prime} - \varPhi^{\prime} \right)\\
- 2\left( 3\Hat{m}^{\prime} + \Hat{\ell}^{\prime} + 6\Hat{m}^{2} + \Hat{\ell}^{2} + 3\Hat{m}\Hat{\ell} \right)\left( \varPsi + \mathcal{X} - \varPhi \right) =
\frac{\chi}{g}\frac{a^{\prime}}{\mathcal{L}}\left[ a^{\prime}\left( \mathcal{X} + \varPhi \right) + \varTheta_{\theta}^{\prime} \right]
- \mu\chi\left( 1 - 2\varSigma_{1}\cos\frac{f}{2} + 2\varSigma_{2}\sin\frac{f}{2} \right)\\
+ k\chi\left[
\frac{f^{\prime2}}{2}\mathcal{X} + \frac{2v}{L^{2}}\varPhi + \frac{v_{a}}{L^{2}}\varTheta_{\theta}
+ f^{\prime}\sin{f}{2}\varSigma_{1}^{\prime} - \frac{2a^{2}}{L^{2}}\cos{f}{2}\varSigma_{1}
- f^{\prime}\cos{f}{2}\varSigma_{2}^{\prime} - \frac{2\left( n - a \right)^{2}}{L^{2}}\sin{f}{2}\varSigma_{2}
\right],
\label{eq:p_ein_mm_s2}
\end{multline}
the $(r,r)$ component is
\begin{multline}
\frac{1}{M^{2}}\square\left( 3\varPsi + \varPhi \right) + 4\left( 3\Hat{m} + \Hat{\ell} \right)\varPsi^{\prime} + 4\Hat{m}^{\prime}\varPhi^{\prime} = 
- \frac{\chi}{g}\frac{a^{\prime}}{L^{2}}\left( a^{\prime}\varPhi + \varTheta_{\theta}^{\prime} \right)
- \mu\chi\left\{ 1 + 2\left[ \mathcal{X}\left( 1 - \cos{f} \right) - \varSigma_{1}\cos\frac{f}{2} + \varSigma_{2}\sin\frac{f}{2} \right] \right\}\\
+ k\chi\left[
\frac{2v}{L^{2}}\left( \varPhi - \mathcal{X} \right) + \frac{v_{a}}{L^{2}}\varTheta_{\theta}
- f^{\prime}\sin\frac{f}{2}\varSigma_{1}^{\prime} - \frac{2a^{2}}{L^{2}}\cos\frac{f}{2}\varSigma_{1}
+ f^{\prime}\cos\frac{f}{2}\varSigma_{2}^{\prime} - \frac{2\left( n - a \right)^{2}}{L^{2}}\sin\frac{f}{2}\varSigma_{2}
\right] - 2\chi\Lambda_{b}\mathcal{X},
\label{eq:p_ein_rr_s2}
\end{multline}
the $(\theta,\theta)$ component is
\begin{multline}
\frac{1}{M^{2}}\square\left( 3\varPsi + \mathcal{X} \right)
+ 4\left[
\varPsi^{\prime\prime} + \Hat{m}\left( 5\varPsi^{\prime} - \mathcal{X}^{\prime} \right)
+ \left( 2\Hat{m}^{\prime} + 5\Hat{m}^{2} \right)\left( \varPhi - \mathcal{X} \right)
\right]\\
= - \frac{\chi}{g}\frac{a^{\prime}}{L^{2}}\left( a^{\prime}\mathcal{X} - \varTheta_{\theta}^{\prime} \right)
- \mu\chi\left\{ 1 + 2\left[ \varPhi\left( 1 - \cos{f} \right) - \varSigma_{1}\cos\frac{f}{2} + \varSigma_{2}\sin\frac{f}{2} \right] \right\}\\
+ k\chi\left[
\frac{f^{\prime2}}{2}\left( \mathcal{X} - \varPhi \right) - \frac{v_{a}}{L^{2}}\varTheta_{\theta}
+ f^{\prime}\sin\frac{f}{2}\varSigma_{1}^{\prime} + \frac{2a^{\prime}}{L^{2}}\cos\frac{f}{2}\varSigma_{1}
- f^{\prime}\cos\frac{f}{2}\varSigma_{2}^{\prime} + \frac{2\left( n - a \right)^{2}}{L^{2}}\sin\frac{f}{2}\varSigma_{2}
\right] - 2\chi\Lambda_{b}\varPhi,
\label{eq:p_ein_tt_s2}
\end{multline}
and the $(\mu,r)$ component is 
\begin{equation}
3\varPsi^{\prime} + \varPhi^{\prime} - \left( \Hat{m} - \Hat{\ell} \right)\varPhi - \left( 3\Hat{m} + \Hat{\ell} \right)\mathcal{X}
= \frac{\chi}{g}\frac{a^{\prime}}{L^{2}}\varTheta_{\theta}
+ k\chi\left( f^{\prime}\sin\frac{f}{2}\varSigma_{1} - f^{\prime}\cos\frac{f}{2}\varSigma_{2} \right).
\label{eq:p_ein_mr_s2}
\end{equation}
Next for the perturbed matter field equations,
the real parts of the perturbed $\mathbb{C}P^{1}$ field equations are
\begin{multline}
\frac{1}{M^{2}}\square\varSigma_{1} + \varSigma_{1}^{\prime\prime}
+ \left(4\Hat{m} + \Hat{\ell}\right)\left(\varSigma_{1}^{\prime} + \mathcal{X}f^{\prime}\sin\frac{f}{2}\right)
+ \mathcal{X}f^{\prime\prime}\sin\frac{f}{2}
\\
- \left(2\varPsi^{\prime} +\frac{\varPhi^{\prime} - \mathcal{X}^{\prime}}{2}\right)f^{\prime}\sin\frac{f}{2}
+ \left(\frac{f^{\prime 2}}{2}\mathcal{X} + \frac{2a^{2}}{L^{2}}\varPhi + \frac{2a}{L^{2}}\varTheta_{\theta}\right)\cos\frac{f}{2}
- \left(\frac{a^{2}}{L^{2}} - \frac{\mu}{k}\right)\varSigma_{1} = 0
\label{eq:p_cp1_1r}
\end{multline}
and
\begin{multline}
\frac{1}{M^{2}}\square\varSigma_{2} + \varSigma_{2}^{\prime\prime}
+ \left(4\Hat{m} + \Hat{\ell}\right)\left(\varSigma_{2}^{\prime} - \mathcal{X}f^{\prime}\cos\frac{f}{2}\right)
- \mathcal{X}f^{\prime\prime}\cos\frac{f}{2}
+ \left(2\varPsi^{\prime} + \frac{\varPhi^{\prime} - \mathcal{X}^{\prime}}{2}\right)f^{\prime}\cos\frac{f}{2}
\\
+ \left[\frac{f^{\prime 2}}{2}\mathcal{X} + \frac{2\left(n - a\right)^{2}}{L^{2}}\varPhi - \frac{2\left(n - a\right)}{L^{2}}\varTheta_{\theta}\right]\sin\frac{f}{2}
- \left[ \frac{\left(n - a\right)^{2}}{L^{2}} + \frac{\mu}{k} \right]\varSigma_{2} = 0,
\label{eq:p_cp1_2r}
\end{multline}
and the only $(\theta)$ component of the perturbed $U(1)$ gauge field equation is
\begin{equation}
\frac{1}{M^{2}}\square\varTheta_{\theta}
+ \varTheta_{\theta}^{\prime\prime}
+ \left( 4\Hat{m} - \Hat{\ell} \right)\varTheta_{\theta}^{\prime}
- a^{\prime}\left( 4\varPsi^{\prime} - \mathcal{X}^{\prime} - \varPhi^{\prime} \right)
+ 2\mathcal{X}v_{a}
= 8kg\left[ \varTheta_{\theta} - 2a\varSigma_{1}\cos\frac{f}{2} + 2\left( n - a \right)\varSigma_{2}\sin\frac{f}{2} \right].
\label{eq:p_u1_t_s2}
\end{equation}
\end{widetext}

The properties of each evolution equations are as follows:
Eq.(\ref{eq:p_ein_mn_s2}) is a constraint for the three fluctuations of the geometry;
also Eqs.(\ref{eq:p_ein_rr_s2}) and (\ref{eq:p_ein_mr_s2}) are constraints for all the fluctuations belonging to the sector II,
and Eqs.(\ref{eq:p_ein_mm_s2}), (\ref{eq:p_ein_tt_s2}), and (\ref{eq:p_cp1_1r})-(\ref{eq:p_u1_t_s2}) are dynamical evolution equations.
Owing to the constraint (\ref{eq:p_ein_mn_s2}) for the geometry fluctuations,
we are able to evaluate about the five degrees of freedom,
\emph{i.e.}, $\varPsi, \mathcal{X}, \varSigma_{1}, \varSigma_{2}$, and $\varTheta_{\theta}$.

If we consider the special situation where all five fluctuations are massless,
we obtain the asymptotic equations at infinity
from Eqs.(\ref{eq:p_ein_rr_s2}), (\ref{eq:p_ein_mr_s2}), and (\ref{eq:p_cp1_1r})-(\ref{eq:p_u1_t_s2}),
\begin{align}
&2\Tilde{\varPsi}^{\prime}(\rho) - \Tilde{\mathcal{X}}^{\prime}(\rho) - 5c\Tilde{\mathcal{X}}(\rho) 
= \frac{\alpha\gamma}{4c}\left( 1 - 2\Tilde{\varSigma}_{1}(\rho) \right),
\notag\\
&\Tilde{\varPsi}^{\prime}(\rho) - \Tilde{\mathcal{X}}^{\prime}(\rho) + 4c\Tilde{\mathcal{X}}(\rho) = 0
\label{eqs:peq_geom_s2_inf}
\end{align}
for the fluctuations of the geometry and
\begin{align}
&\Tilde{\varTheta}_{\theta}^{\prime\prime}(\rho) - 3c\Tilde{\varTheta}_{\theta}^{\prime}(\rho) = 8\Tilde{\varTheta}_{\theta}(\rho),
\notag\\
&\Tilde{\varSigma}_{1}^{\prime\prime}(\rho) - 5c\Tilde{\varSigma}_{1}^{\prime}(\rho) + \gamma\Tilde{\varSigma}_{1}(\rho) = 0,
\notag\\
&\Tilde{\varSigma}_{2}^{\prime\prime}(\rho) - 5c\Tilde{\varSigma}_{2}^{\prime}(\rho) - \gamma\Tilde{\varSigma}_{2}(\rho) = 0
\label{eqs:peq_matters_s2_inf}
\end{align}
for the fluctuations of the matter fields,
where we again use the variables rescaled with respect to $\rho$. 
Consider the fluctuation of the constraint for the $\mathbb{C}P^{1}$ doublet $|\bm{z}|^{2} = 1$,
that is,
\begin{align}
\delta|\bm{z}|^{2} &= \delta\left( z^{*a}z^{a} \right) = \delta z^{*a}z^{a} + z^{*a}\delta z^{a} 
\notag\\
&= \Tilde{\varSigma}_{1}\cos\frac{\Tilde{f}}{2} + \Tilde{\varSigma}_{2}\sin\frac{\Tilde{f}}{2} = \delta(1) = 0,
\notag
\end{align}
we find the constraint for the perturbed $\mathbb{C}P^{1}$ fields
\begin{equation}
\Tilde{\varSigma}_{1}(\rho) = \Tilde{\varSigma}_{2}(\rho)\tan\frac{\Tilde{f}}{2}.
\notag
\end{equation}
$\Tilde{\varSigma}_{1}(\rho)$ goes to zero at infinity
since it must hold the boundary conditions for the backgrounds (\ref{eqs:matter_bc}).
Hence we determine the asymptotic solutions for Eqs.(\ref{eqs:peq_matters_s2_inf})
\begin{align}
&\Tilde{\varTheta}_{\theta}(\rho) \propto \exp\left[ \frac{3c}{2}\left( 1 \pm \sqrt{1 + \frac{32}{9c^{2}}} \right)\rho \right],
\notag\\
&\Tilde{\varSigma}_{1}(\rho) \simeq 0,
\notag\\
&\Tilde{\varSigma}_{2}(\rho) \propto \exp\left[ \frac{5c}{2}\left( 1 \pm \sqrt{1 + \frac{8\gamma}{25c^{2}}} \right)\rho \right].
\label{eqs:asym_inf_matter_s2}
\end{align}
Also,
we get the asymptotic solutions of the fluctuations for the geometry from the above results and Eqs.(\ref{eq:p_ein_mn_s2}) and (\ref{eqs:peq_geom_s2_inf}):
\begin{align}
&\Tilde{\mathcal{X}}(\rho) = \frac{\alpha\gamma}{4c} + \Tilde{\chi}^{(\infty)}e^{13c\rho},
\notag\\
&\Tilde{\varPsi}(\rho) = \Tilde{\psi}^{(\infty)} - \alpha\gamma\rho + \frac{9}{13}\Tilde{\chi}^{(\infty)}e^{13c\rho},
\notag\\
&\Tilde{\varPhi}(\rho) = \Tilde{\phi}^{(\infty)} - \frac{\alpha\gamma}{4c} + 2\alpha\gamma\rho - \frac{31}{13}\Tilde{\chi}^{(\infty)}e^{13c\rho},
\label{eqs:asym_inf_geom_s2}
\end{align}
where $\Tilde{\chi}^{(\infty)}, \Tilde{\psi}^{(\infty)}$ and $\Tilde{\phi}^{(\infty)}$ are arbitrary integral constants
which satisfy Eq.(\ref{eq:p_ein_mn_s2}).
Eqs.(\ref{eqs:asym_inf_matter_s2}) and (\ref{eqs:asym_inf_geom_s2}) clearly show that
the three scalar mode fluctuations of the geometry diverge at infinity
even though the other three fluctuations of the matters are regular. 

On the other hand,
we investigate asymptotic behaviors at the origin as the same method in Sec.{\ref{sc:as}}.
Consequently we get the asymptotic solutions as follows:
\begin{align}
&\Tilde{\varPsi}(\rho) \simeq - \frac{\alpha}{8}\left( \gamma + 4\mathcal{B}\Tilde{\vartheta}_{\theta}^{(0)} \right)\rho^{2} + \mathcal{O}(\rho^{3}),
\notag\\
&\Tilde{\mathcal{X}}(\rho) \simeq \Tilde{\chi}^{(0)}\rho^{2} + \mathcal{O}(\rho^{3}),
\notag\\
&\Tilde{\varTheta}_{\theta}(\rho) \simeq \Tilde{\vartheta}_{\theta}^{(0)}\rho^{2} + \mathcal{O}(\rho^{3}),
\notag\\
&\Tilde{\varSigma}_{1}(\rho) \simeq \Tilde{\sigma}_{1}^{(0)}\rho^{3} + \mathcal{O}(\rho^{4}),
\notag\\
&\Tilde{\varSigma}_{2}(\rho) \simeq \frac{\mathcal{A}}{2}\Tilde{\sigma}_{1}^{(0)}\rho^{4} + \mathcal{O}(\rho^{5}),
\label{eqs:asym_origin_s2}
\end{align}
where $\Tilde{\vartheta}_{\theta}^{(0)}$, $\Tilde{\chi}^{(0)}$, and $\Tilde{\sigma}_{1}^{(0)}$ are arbitrary constants.
For the rest of degrees $\Tilde{\varPhi}(\rho)$,
we similarly evaluate by plugging the asymptotic solutions into the constraint (\ref{eq:p_ein_mn_s2}).

Therefore, from Eqs.(\ref{eqs:asym_inf_matter_s2})-(\ref{eqs:asym_origin_s2}),
we conclude that the geometry parts of the fluctuations belonging to the sector II are not normalizable degrees
whereas the matter parts of them are potentially normalizable ones.
Of course,
we should be practically required some numerical calculations in order to find exact solutions for the evolution equations.
 
\section{Conclusion}\label{sc:cncl}
In this paper we have investigated a new brane model in six dimensions, 
constructed the thick brane solution by the Maxwell gauged $\mathbb{C}P^{1}$ model.
The origin of this model is naturally obtained by projecting $O(3)$ sigma model onto a complex space
and exchanging a $U(1)$ gauge term with the native composite connection.
As we have shown,
our model can realize localizing gravity around the 3-brane with certain parameter space of the model
and attain a finite four-dimensional Planck mass (Eq.(\ref{eq:planckmass})).
Our results clearly suggest the possibility of a new variety of braneworlds based on 
classical field theory because there are many variants for $O(3)$ sigma model
{\cite{Piette:1994xf,Arthur:1996uu,Tripathy:1998sg,Piette:1994jt,Tchrakian:1995ji,
Gladikowski:1995sc,Kimm:1997bi,Keeffe:1998kn,Otsu:2003fq,Otsu:2005mf}} (see also Ref.{\cite{Kodama:2008zz}}).
Besides, relevant previous works are almost constructed from the classical solutions in gauge theory;
in particular Abelian-Higgs (AH) vortex in six dimensions {\cite{Giovannini:2001hh}}.
By using the different model from AH vortex,
we have showed that a solitonic nature of vortices can attain gravity localization around the brane core in codimension-2 braneworld models.
Since such mechanism using any topological soliton models does not always work in higher codimensional models {\cite{Gherghetta:2000jf}}, 
it is significant that our model practically obtain similar results to AH vortex .

Our another aim at the present paper is to analyze the linear stability of the model,
for all fluctuating fields around the background classical solution.
We have concentrated to study only their zero modes here.
For the fluctuations of the geometry,
the tensor zero mode is always localized around the origin, 
on the other hand neither the vector nor scalar modes are localized.
The fluctuations of the matter fields are potentially localized since they are regular at infinity and near the origin.
Here, ``Potentially'' means that we could not find any analytical solutions.
In order to determine whether these fluctuations of matters are true stable or not,
we need vast numerical calculations for all the perturbed equations of motion presented in Appendix {\ref{ap:pfes}}. 
We therefore conclude the brane described by Maxwell $\mathbb{C}P^{1}$ model is a quasi-stable configuration. 

For the linear stability analysis on the zero mode fluctuations,
other thick brane models in six dimensions
{\cite{Giovannini:2002mk,Giovannini:2002sb,RandjbarDaemi:2002pq}}
have obtained similar results to us.
These codimension-2 models including ours have a serious drawback as pointed out in Ref.{\cite{Peter:2003zg}}.
One of aims in our paper is to study such problem by using other model with different topology from AH vortex,
but we finally encounter a similar difficulty.
However, we believe that thick branes include rich physical implications than thin ones,
thus we should tackle with the problem. 
In recent years, the many attempts to resolve the problem of the quasi-stability by introducing other degrees of freedom.
We are also currently studying such quasi-stability problem about our solutions and the results will be reported in near future. 
 

\begin{acknowledgments}
We deeply appreciate to Noriko Shiiki for many useful advices and comments. 
This work is motivated by earlier project she initiated.
\end{acknowledgments}

\begin{widetext}
\appendix
\section{Longitudinal gauge system}\label{ap:gifl}
Mathematically,
the problem of describing evolution equations for small perturbations in general relativity
is equivalent for solving the Einstein equation linearized about an expanding background metric tensor.
This procedure is straightforward but
there are complicated issues concerning the freedom of gauge,
\emph{i.e.}, the choice of background coordinates.
In this appendix we derive the gauge-invariant fluctuation for all fields (\ref{eq:gi_pmet})-(\ref{eqs:gi_pmatter}) in Sec.{\ref{sc:lgp}}
by imposing general fluctuations upon the longitudinal gauge conditions.

First of all,
we introduce the first-order general fluctuation of the background metric tensor (\ref{eq:metric}) given by
\begin{equation}
\delta\mathcal{G}_{MN}
= \delta^{(t)}\mathcal{G}_{MN} + \delta^{(v)}\mathcal{G}_{MN} + \delta^{(s)}\mathcal{G}_{MN}
=\begin{pmatrix}
2M^{2}\mathcal{U}_{\mu\nu}&M\mathcal{V}_{\nu}&ML\mathcal{W}_{\nu} \cr
M\mathcal{V}_{\mu}&2\chi &L\omega \cr
ML\mathcal{W}_{\mu}&L\omega &2L^{2}\phi \cr
\end{pmatrix},
\label{eq:fluc}
\end{equation}
where
\begin{equation}
\mathcal{U}_{\mu\nu} = h_{\mu\nu} + \partial_{(\mu}\mathcal{I}_{\nu)} + \eta_{\mu\nu}\psi + \partial_{\mu}\partial_{\nu}\mathcal{O};\;\;\;
\mathcal{V}_{\mu} = \mathcal{J}_{\mu} + \partial_{\mu}\mathcal{P},\;\;\;
\mathcal{W}_{\mu} = \mathcal{K}_{\mu} + \partial_{\mu}\mathcal{Q},
\notag
\end{equation}
with the divergenceless and the traceless constraints
\begin{equation}
\partial_{\mu}h^{\mu}_{\nu} = 0,\;\;\;h_{\mu}^{\mu} = 0;\;\;\;
\partial_{\mu}\mathcal{I}^{\mu} = 0,\;\;\;\partial_{\mu}\mathcal{J}^{\mu} = 0,\;\;\;\partial_{\mu}\mathcal{K}^{\mu} = 0.
\label{eqs:g_condition}
\end{equation}
In above fluctuations,
$h_{\mu\nu}$ is a tensor mode,
$\mathcal{I}_{\mu}, \mathcal{J}_{\mu}$, and $\mathcal{K}_{\mu}$ are vector modes,
and finally $\psi, \mathcal{O}, \mathcal{P}, \mathcal{Q}, \chi, \phi$, and $\omega$ are scalar modes,
which depend on all the spatial coordinates $x^{M} = (x^{\mu}, r, \theta)$.

As we mentioned before,
any gauge-invariant quantities are required to be invariant with respect to the choice of background coordinates.
Here an infinitesimal coordinates transformations are defined by
\begin{equation}
x^{M} \to \Tilde{x}^{M} = x^{M} + \xi^{M}(x^{N})
\label{eq:inf_trans}
\end{equation}
with
\begin{equation}
\xi^{M}(x^{N}) = \left( M^{2}\epsilon^{\mu},\epsilon^{r},L^{2}\epsilon^{\theta} \right),
\notag
\end{equation}
where the local infinitesimal parameter $\epsilon^{M}$ depends on $x^{M}$.
Also we take the four-dimensional sector of the transformations as follows:
\begin{equation}
\epsilon_{\mu} = \partial_{\mu}\epsilon + \zeta_{\mu}.
\label{eq:gf_trans}
\end{equation}
If the perturbed metric tensor (\ref{eq:fluc}) is transformed $\delta\Tilde{\mathcal{G}}_{MN}$ 
under the transformations (\ref{eq:inf_trans}), which can be written by
\begin{equation}
\delta\Tilde{\mathcal{G}}_{MN} = \delta \mathcal{G}_{MN} - \pounds_{\xi}\mathcal{G}_{MN}.
\label{eq:after_gt}
\end{equation}
where $\pounds_{\xi}$ is the Lie derivative defined as
\begin{equation}
\pounds_{\xi}\mathcal{G}_{MN}
= \xi^{L}\partial_{L}\mathcal{G}_{MN} + \mathcal{G}_{ML}\partial_{N}\xi^{L} + \mathcal{G}_{LN}\partial_{M}\xi^{L}
= \nabla_{M}\xi_{N} + \nabla_{N}\xi_{M},
\notag
\end{equation}
where $\nabla_{M}$ means the covariant derivative with respect to the metric tensor.
When one calculate the Lie covariant derivative involved in the perturbed quantities and others,
one may use the original metric tensor (\ref{eq:metric}) and the original Christoffel connections
\begin{equation}
\Gamma^{\mu}_{\alpha r} = \frac{M^{\prime}}{M}\delta_{\alpha}^{\mu},\;\;\;
\Gamma^{\theta}_{r\theta} = \frac{L^{\prime}}{L},\;\;\;
\Gamma^{r}_{\alpha\beta} = - MM^{\prime}\eta_{\alpha\beta},\;\;\;
\Gamma^{r}_{\theta\theta} = -LL^{\prime}.
\notag
\end{equation} 
Inserting the perturbed metric tensor (\ref{eq:fluc}) into Eq.(\ref{eq:after_gt}),
we obtain the explicit forms of $\delta\Tilde{\mathcal{G}}_{MN}$.
Taking some linear combination of them,
we can obtain following forms of the scalar functions
\begin{align}
&\Tilde{\varPsi} := \Tilde{\psi} - M^{\prime}\left( \Tilde{\mathcal{P}} - M\Tilde{\mathcal{O}}^{\prime} \right),
\notag\\
&\Tilde{\mathcal{X}} := \Tilde{\chi} - \left[ M\left( \Tilde{\mathcal{P}} - M\Tilde{\mathcal{O}}^{\prime} \right) \right]^{\prime},
\notag\\
&\Tilde{\varPhi} := \Tilde{\phi} - \frac{ML^{\prime}}{L}\left( \Tilde{\mathcal{P}} - M\Tilde{\mathcal{O}}^{\prime} \right)
- \left[ \frac{M}{L}\left( \Tilde{\mathcal{Q}} - \frac{M}{L}\dot{\Tilde{\mathcal{O}}} \right) \right]^{\cdot},
\notag\\
&\Tilde{\varOmega} := \Tilde{\omega} - \frac{1}{L}\left[ M\left( \Tilde{\mathcal{P}} - M\Tilde{\mathcal{O}}^{\prime} \right) \right]^{\cdot}
- L\left[ \frac{M}{L}\left( \Tilde{\mathcal{Q}} - \frac{M}{L}\dot{\Tilde{\mathcal{O}}} \right) \right]^{\prime},
\label{eqs:gi_s_g}
\end{align}
where the prime and the over-dot denote the derivative with respect to the bulk radius $r$ and the bulk angle $\theta$, respectively.
Similarly two gauge-invariant vector functions
\begin{equation}
\Tilde{\mathcal{E}}_{\mu} := \Tilde{\mathcal{J}}_{\mu} - M\Tilde{\mathcal{I}}_{\mu}^{\prime},\;\;\;
\Tilde{\mathcal{F}}_{\mu} := \Tilde{\mathcal{K}}_{\mu} - \frac{M}{L}\dot{\Tilde{\mathcal{I}}}_{\mu}
\label{eqs:gi_v_g}
\end{equation}
and one gauge-invariant tensor function
\begin{equation}
\Tilde{h}_{\mu\nu} = h_{\mu\nu}
\label{eq:gi_t_g}
\end{equation}
are obtained.

On the other hand, for the matter fields,
the original $\mathbb{C}P^{1}$ and $U(1)$ gauge fields (\ref{eqs:matter_ansatz}) are fluctuated by
\begin{equation}
\delta \bm{z} = (\delta z_{1}, \delta z_{2})^{T} = (\sigma_{1}(x^{M})e^{-in_{a}\theta}, \sigma_{2}(x^{M}))^{T};\;\;\;
\delta A_{M} = (\vartheta_{\mu}(x^{N}), \vartheta_{r}(x^{N}), \vartheta_{\theta}(x^{N}))
\label{eqs:p_matter}
\end{equation}
with the decoupled vector component
\begin{equation}
\vartheta_{\mu} = \varTheta_{\mu} + \partial_{\mu}\vartheta,
\notag
\end{equation}
where $\varTheta_{\mu}$ and $\vartheta$ are a vector and a scalar fluctuation for the $U(1)$ gauge field, respectively.
We obtain transforms of the matter fields under Eq.(\ref{eq:inf_trans}) in the same way as the geometry.
Hence we can defined six gauge-invariant scalar functions given by
\begin{align}
\Tilde{\varSigma}_{1} &:= \Tilde{\sigma}_{1} + \frac{f^{\prime}}{2}M\left( \Tilde{\mathcal{P}} - M\Tilde{\mathcal{O}}^{\prime} \right)\sin\frac{f}{2}
+ in_{a}\frac{M}{L}\left( \Tilde{\mathcal{Q}} - \frac{M}{L}\dot{\Tilde{\mathcal{O}}} \right)\cos\frac{f}{2},
\notag\\
\Tilde{\varSigma}_{2} &:= \Tilde{\sigma}_{2} - \frac{f^{\prime}}{2}M\left( \Tilde{\mathcal{P}} - M\Tilde{\mathcal{O}}^{\prime} \right)\cos\frac{f}{2}
\label{eqs:gi_cp1}
\end{align}
for the $\mathbb{C}P^{1}$ field and
\begin{align}
\Tilde{\varTheta}_{\mu} &= \varTheta_{\mu},\;\;\;
\Tilde{\varTheta} := \Tilde{\vartheta} - A_{\theta}\frac{M}{L}\left( \Tilde{\mathcal{Q}} - \frac{M}{L}\dot{\Tilde{\mathcal{O}}} \right),
\notag\\
\Tilde{\varTheta}_{r}
&:= \Tilde{\vartheta}_{r} - A_{\theta}\left[ \frac{M}{L}\left( \Tilde{\mathcal{Q}} - \frac{M}{L}\dot{\Tilde{\mathcal{O}}} \right) \right]^{\prime}
- A_{\theta}\frac{ML^{\prime}}{L^{2}}\left( \Tilde{\mathcal{Q}} - \frac{M}{L}\dot{\Tilde{\mathcal{O}}} \right),
\notag\\
\Tilde{\varTheta}_{\theta}
&:= \Tilde{\vartheta}_{\theta} - A_{\theta}\left[ \frac{M}{L}\left( \Tilde{\mathcal{Q}} - \frac{M}{L}\dot{\Tilde{\mathcal{O}}} \right) \right]^{\cdot}
- A_{\theta}^{\prime}M\left( \Tilde{\mathcal{P}} - M\Tilde{\mathcal{O}}^{\prime} \right)
\label{eqs:gi_gauge}
\end{align}
for the gauge field where the vector mode $\Tilde{\varTheta}_{\mu}$is automatically gauge-invariant variable like $\Tilde{h}_{\mu\nu}$.

Finally we take the longitudinal (or Conformal Newtonian) gauge conditions {\cite{Mukhanov:1990me}},
where $\Tilde{\mathcal{O}}, \Tilde{\mathcal{P}}, \Tilde{\mathcal{Q}}, \Tilde{\mathcal{I}}_{\mu}$ are to be zero
after the infinitesimal coordinates transformations (\ref{eq:inf_trans}).
These gauges can be written in terms of $\mathcal{O},\mathcal{P},\mathcal{Q},\mathcal{I}_{\mu}$ as
\begin{equation}
\epsilon = \mathcal{O},\;\;\;
\epsilon_{r} = M\left( \mathcal{P} - M\mathcal{O}^{\prime} \right),\;\;\;
\epsilon_{\theta} = \frac{M}{L}\left( \mathcal{Q} - \frac{M}{L}\dot{\mathcal{O}} \right),
\notag
\end{equation}
and $\zeta_{\mu} = \mathcal{I}_{\mu}$.
Then the gauge-invariant fluctuations of the geometry (\ref{eqs:gi_s_g})-(\ref{eq:gi_t_g})
and of the matter fields (\ref{eqs:gi_cp1}) and (\ref{eqs:gi_gauge}) are exactly equivalent to the original one under the transformations (\ref{eq:inf_trans}).
In our perturbative calculation,
therefore the independent gauge degrees of freedom are 
$
\Tilde{\varPsi},\;
\Tilde{\mathcal{X}},\;
\Tilde{\varPhi},\;
\Tilde{\varOmega},\;
\Tilde{\mathcal{E}}_{\mu},\;
\Tilde{\mathcal{F}}_{\mu},\;
\Tilde{h}_{\mu\nu},\;
\Tilde{\varSigma}_{1},\;
\Tilde{\varSigma}_{2},\;
\Tilde{\varTheta}_{\mu},\;
\Tilde{\varTheta},\;
\Tilde{\varTheta}_{r}
$, and $\Tilde{\varTheta}_{\theta}$.
Also we omit the tilde for convenience in Sec.{\ref{sc:lgp}}. 

\section{Perturbed quantities}\label{ap:pfes}
The explicit evaluation of the perturbed equations of motion (\ref{eqs:per_eqs}) is proceeded here.
We shall show the clear forms of the perturbed Christoffel connection, Einstein tensor, energy-momentum tensor,
and equations for the matter fields in the below. 

\subsection{Christoffel connection}
Inserting the explicit form of the gauge-invariant perturbed metric tensor (\ref{eq:gi_pmet}) into Eq.(\ref{eq:p_christ}),
one can get all components of the perturbed Christoffel connection after lengthly calculations.
The $(\mu,M,N)$ components are
\begin{align}
&\delta\Gamma^{\mu}_{\alpha\beta}
= 2\partial_{(\alpha}\mathcal{U}_{\beta)}^{\mu}
- \partial^{\mu}\mathcal{U}_{\alpha\beta}
+ M^{\prime}\eta_{\alpha\beta}\mathcal{E}^{\mu},
&&\delta\Gamma^{\mu}_{\alpha r}
= \mathcal{U}_{\alpha}^{\mu\prime}
+ \frac{1}{2M} \left(
\partial_{\alpha}\mathcal{E}^{\mu} - \partial^{\mu}\mathcal{E}_{\alpha}
\right),
\notag\\
&\delta\Gamma^{\mu}_{\alpha\theta}
= \dot{\mathcal{U}}_{\alpha}^{\mu}
+ \frac{L}{2M} \left(
\partial_{\alpha}\mathcal{F}^{\mu} - \partial^{\mu}\mathcal{F}_{\alpha}
\right),
&&\delta\Gamma^{\mu}_{rr}
= \frac{1}{M} \left( \mathcal{E}^{\mu\prime} + \Hat{m}\mathcal{E}^{\mu} \right)
- \frac{1}{M^{2}}\partial^{\mu}\mathcal{X},
\notag\\
&\delta\Gamma^{\mu}_{r\theta}
= \frac{1}{2M}\dot{\mathcal{E}}^{\mu}
+ \frac{L}{2M} \left[
\mathcal{F}^{\mu\prime} + \left(\Hat{m} - \Hat{\ell}\right)\mathcal{F}^{\mu}
\right]
- \frac{L}{2M^{2}}\partial^{\mu}\varOmega,
&&\delta\Gamma^{\mu}_{\theta\theta}
= \frac{L}{M} \left( L^{\prime}\mathcal{E}^{\mu} + \Dot{\mathcal{F}}^{\mu} \right)
- \frac{L^{2}}{M^{2}}\partial^{\mu}\varPhi, 
\end{align}
the $(r,M,N)$ components are
\begin{align}
&\delta\Gamma^{r}_{\alpha\beta}
= - M^{2}\left( \mathcal{U}_{\alpha\beta}^{\prime} + 2\Hat{m}\mathcal{U}_{\alpha\beta} \right)
+ M\partial_{(\alpha}\mathcal{E}_{\beta)}
+ 2M^{2}\eta_{\alpha\beta}\Hat{m}\mathcal{X},
&&\delta\Gamma^{r}_{\alpha r}
= - M^{\prime}\mathcal{E}_{\alpha} + \partial_{\alpha}\mathcal{X},
\notag\\
&\delta\Gamma^{r}_{\alpha\theta}
= \frac{M}{2}\Dot{\mathcal{E}}_{\alpha}
- \frac{ML}{2}\left[
\mathcal{F}_{\alpha}^{\prime} + \left(\Hat{m} + \Hat{\ell}\right)\mathcal{F}_{\alpha}
\right]
+ \frac{L}{2}\partial_{\alpha}\varOmega,
&&\delta\Gamma^{r}_{rr} = \mathcal{X}^{\prime},
\notag\\
&\delta\Gamma^{r}_{r\theta} = \Dot{\mathcal{X}} - L^{\prime}\varOmega,
&&\delta\Gamma^{r}_{\theta\theta} = L\Dot{\varOmega}
- L^{2}\left[ \varPhi^{\prime} + 2\Hat{\ell}\left( \varPhi - \mathcal{X} \right)\right], 
\end{align}
and the $(\theta,M,N)$ components are
\begin{align}
&\delta\Gamma^{\theta}_{\alpha\beta}
= - \frac{M^{2}}{L^{2}}\Dot{\mathcal{U}}_{\alpha\beta} + \frac{M}{L}\partial_{(\alpha}\mathcal{F}_{\beta)}
+ \frac{M}{L}\eta_{\alpha\beta}M^{\prime}\varOmega,
&&\delta\Gamma^{\theta}_{\alpha r}
= - \frac{M}{2L^{2}}\Dot{\mathcal{E}}_{\alpha}
+ \frac{M}{2L}\left[ \mathcal{F}_{\alpha}^{\prime} - \left(\Hat{m} - \Hat{\ell}\right)\mathcal{F}_{\alpha} \right]
+ \frac{1}{2L}\partial_{\alpha}\varOmega,
\notag\\
&\delta\Gamma^{\theta}_{\alpha \theta} = \partial_{\alpha}\varPhi,
&&\delta\Gamma^{\theta}_{rr}
= - \frac{1}{L^{2}}\Dot{\mathcal{X}} + \frac{1}{L}\left( \varOmega^{\prime} + \Hat{\ell}\varOmega \right),
\notag\\
&\delta\Gamma^{\theta}_{r\theta} = \varPhi^{\prime},
&&\delta\Gamma^{\theta}_{\theta\theta} = \Dot{\varPhi} + L^{\prime}\varOmega.
\end{align}

\subsection{Einstein tensor}
Using the explicit forms of the perturbed Christoffel connections which we obtained,
one can also evaluate the perturbed Riemann tensor (\ref{eq:p_riemann}) and Ricci scalar (\ref{eq:p_rscalar}).
Inserting these perturbed quantities into Eq.(\ref{eq:p_etensor}),
one can achieve six components of the perturbed Einstein tensor.
The four-dimensional component $(\mu,\nu)$
\begin{align}
\delta G_{\mu\nu}
&= - \square h_{\mu\nu} - \frac{M^{2}}{L^{2}} \Ddot{h}_{\mu\nu}
- M^{2}\left[
h_{\mu\nu}^{\prime\prime}
+ \left( 4\Hat{m} + \Hat{\ell} \right)h_{\mu\nu}^{\prime}
- 2h_{\mu\nu} \left( 3\Hat{m}^{\prime} + \Hat{\ell}^{\prime} + 6\Hat{m}^{2} + \Hat{\ell}^{2} + 3\Hat{m}\Hat{\ell} \right)
\right]
\notag\\
&+ M\left[
\partial_{(\mu}\mathcal{E}_{\nu)}^{\prime} + \left( 3\Hat{m} + \Hat{\ell} \right) \partial_{(\mu}\mathcal{E}_{\nu)}
\right]
+ \frac{M}{L}\partial_{(\mu}\Dot{\mathcal{F}}_{\nu)}
- \left( \partial_{\mu}\partial_{\nu} - \eta_{\mu\nu}\square \right) \left( 2\varPsi + \mathcal{X} + \varPhi \right)
\notag\\
&+ M^{2} \eta_{\mu\nu} \Bigg\{
\frac{1}{L^{2}} \left( 3\Ddot{\varPsi} + \Ddot{\mathcal{X}} \right)
+ 3\varPsi^{\prime\prime} + \varPhi^{\prime\prime}
+ 3 \left( 4\Hat{m} + \Hat{\ell} \right) \varPsi^{\prime}
- \left( 3\Hat{m} + \Hat{\ell} \right) \left( \mathcal{X}^{\prime} - \varPhi^{\prime} \right)
\notag\\
&- 2 \left( 3\Hat{m}^{\prime} + \Hat{\ell}^{\prime} + 6\Hat{m}^{2} + \Hat{\ell}^{2} + 3\Hat{m}\Hat{\ell} \right)
\left( \mathcal{X} - \varPhi \right) - \frac{1}{L}\left[ \Dot{\varOmega}^{\prime} + \left( 3\Hat{m} + \Hat{\ell} \right) \Dot{\varOmega} \right]
\Bigg\}
\label{eq:peeq_munu}
\end{align}
has all mode fluctuations of the perturbed quantities.
The three extra-dimensional components $(r,r)$, $(\theta,\theta)$, and $(r,\theta)$
\begin{align}
\delta G_{rr}
&= \frac{1}{M^{2}} \square\left( 3\varPsi + \varPhi \right)
+ \frac{4}{L} \left( \frac{1}{L}\Ddot{\varPsi} - \Hat{m}\Dot{\varOmega} \right)
+ 4 \left( 3\Hat{m} + \Hat{\ell} \right) \varPsi^{\prime} + 4\Hat{m}\varPhi^{\prime},
\label{eq:peeq_rr}\\
\delta G_{\theta\theta}
&= \frac{L^{2}}{M^{2}} \square\left( 3\varPsi + \mathcal{X} \right)
+ L^{2} \left[
4\varPsi^{\prime\prime}
+ 4\Hat{m} \left( 5\varPsi^{\prime} - \mathcal{X}^{\prime} \right)
+ 4 \left( 2\Hat{m}^{\prime} + 5\Hat{m}^{2} \right) \left( \varPhi - \mathcal{X} \right)
\right],
\label{eq:peeq_thetatheta}\\
\delta G_{r\theta}
&= - \frac{L}{2M^{2}} \square\varOmega + 2L\left( 2\Hat{m}^{\prime} + 5\Hat{m}^{2} \right) \varOmega
- 4 \left[ \Dot{\varPsi}^{\prime} + \left( \Hat{m} - \Hat{\ell} \right) \Dot{\varPsi} - \Hat{m}\Dot{\mathcal{X}} \right]
\label{eq:peeq_rtheta}
\end{align}
have the scalar mode fluctuations only.
Finally, the rest of component $(\mu,r)$ and $(\mu,\theta)$
\begin{align}
\delta G_{\mu r}
&= - \frac{1}{2M} \square\mathcal{E}_{\mu} - \frac{M}{2L^{2}} \Ddot{\mathcal{E}}_{\mu}
+ M\left( 3\Hat{m}^{\prime} + \Hat{\ell}^{\prime} + 6\Hat{m}^{2} + \Hat{\ell}^{2} + 3\Hat{m}\Hat{\ell} \right) \mathcal{E}_{\mu}
+ \frac{M}{2L} \left[ \Dot{\mathcal{F}}_{\mu}^{\prime} - \left( \Hat{m} - \Hat{\ell} \right) \Dot{\mathcal{F}}_{\mu} \right]
\notag\\
&- \partial_{\mu} \left[
3\varPsi^{\prime} + \varPhi^{\prime} - \left( \Hat{m} - \Hat{\ell} \right) \varPhi - \left( 3\Hat{m} + \Hat{\ell} \right) \mathcal{X}
- \frac{1}{2L} \Dot{\varOmega}
\right]
\label{eq:peeq_mur}\\
\delta G_{\mu\theta}
&= - \frac{L}{2M} \square\mathcal{F}_{\mu}
+ \frac{M}{2} \left[ \Dot{\mathcal{E}}_{\mu}^{\prime} + \left( 5\Hat{m} - \Hat{\ell} \right) \Dot{\mathcal{E}}_{\mu} \right]
- \frac{ML}{2}\left[
\mathcal{F}_{\mu}^{\prime\prime} + \left( 4\Hat{m} + \Hat{\ell} \right) \mathcal{F}_{\mu}^{\prime}
- \left( 7\Hat{m}^{\prime} + \Hat{\ell}^{\prime} + 17\Hat{m}^{2} + 2\Hat{\ell}^{2} + \Hat{m}\Hat{\ell} \right) \mathcal{F}_{\mu}
\right]
\notag\\
&+ \partial_{\mu}\left\{
\frac{L}{2} \left[ \varOmega^{\prime} + 2 \left( \Hat{m} + \Hat{\ell} \right) \varOmega \right] - 3\Dot{\varPsi} - \Dot{\mathcal{X}}
\right\}
\label{eq:peeq_mutheta}
\end{align}
have the vector and scalar mode fluctuations.

\subsection{Energy-momentum tensor}
In order to evaluate the perturbed energy-momentum tensor (\ref{eq:gp_emt}),
one need the explicit form of the perturbed matter Lagrangian density (\ref{eq:pb_lagrangian}).
Calculating all the fluctuation terms one can obtain
\begin{multline}
\delta \mathcal{L}_{\mathrm{brane}} =
k \left[ \frac{f^{\prime 2}}{2}\mathcal{X} + \frac{v_{a}}{L^{2}}\varTheta_{\theta} + \frac{2v}{L^{2}}\varPhi
+ f^{\prime}\sin\frac{f}{2}\varSigma_{1}^{\prime} - 2\frac{a^{2}}{L^{2}}\cos\frac{f}{2}\varSigma_{1}
- f^{\prime}\cos\frac{f}{2}\varSigma_{2}^{\prime} - 2\frac{\left( n_{a} - a \right)^{2}}{L^{2}}\sin\frac{f}{2}\varSigma_{2} \right]
\\
+ \frac{1}{g}\frac{a^{\prime}}{L^{2}}
\left[ a^{\prime} \left( \mathcal{X} + \varPhi \right) + \varTheta_{\theta}^{\prime} - \Dot{\varTheta_{r}} \right]
- \mu \left( 1 - 2 \varSigma_{1} \cos \frac{f}{2} + 2 \varSigma_{2} \sin \frac{f}{2} \right).
\label{eq:explicit_pblag}
\end{multline}
Hence all components of the perturbed energy-momentum tensor can be given as follows.
The four-dimensional diagonal component $(\mu,\nu)$
\begin{multline}
\delta T_{\mu\nu} =
- M^{2}\eta_{\mu\nu}\Bigg\{
\frac{1}{g}\frac{a^{\prime}}{L^{2}}\left[ a^{\prime}\left( \varPsi - \mathcal{X} - \varPhi \right) + \Dot{\varTheta}_{r} - \varTheta_{\theta}^{\prime} \right]
+ \mu\left[ 1 + 2\varPsi\left( 1 - \cos{f} \right) - 2\varSigma_{1}\cos\frac{f}{2} + 2\varSigma_{2}\sin\frac{f}{2} \right]
\\
+ k\Big[ \frac{f^{\prime2}}{2}\left( \varPsi - \mathcal{X} \right) + \frac{2v}{L^{2}}\left( \varPsi - \varPhi \right) - \frac{v_{a}}{L^{2}}\varTheta_{\theta}
- f^{\prime}\sin\frac{f}{2}\varSigma_{1}^{\prime}
+ 2\frac{a^{2}}{L^{2}}\cos\frac{f}{2}\varSigma_{1} + f^{\prime}\cos\frac{f}{2}\varSigma_{2}^{\prime}
+ 2\frac{\left( n_{a} - a \right)^{2}}{L^{2}}\sin\frac{f}{2}\varSigma_{2} \Big] \Bigg\}
\\
- 2M^{2}h_{\mu\nu}\left[ k\left( \frac{f^{\prime2}}{4} + \frac{v}{L^{2}} \right) + \frac{1}{g}\frac{a^{\prime2}}{2L^{2}} + \mu\left( 1 - \cos{f} \right) \right]
\label{eq:pemt_munu}
\end{multline}
has the tensor and the scalar mode fluctuations,
even though the corresponding component of the perturbed Einstein tensor contains the all mode.
The two extra-dimensional diagonal components $(r,r)$ and $(\theta,\theta)$
\begin{align}
\delta T_{rr} &=
- \frac{1}{g}\frac{a^{\prime}}{L^{2}}\left( a^{\prime}\varPhi + \varTheta_{\theta}^{\prime} - \Dot{\varTheta_{r}} \right)
- \mu\left[ 1 + 2\mathcal{X}\left( 1 - \cos f \right) - 2\varSigma_{1}\cos\frac{f}{2} + 2\varSigma_{2}\sin\frac{f}{2} \right]
\notag\\
&+k\left[
\frac{2v}{L^{2}}\left( \varPhi - \mathcal{X} \right) + \frac{v_{a}}{L^{2}}\varTheta_{\theta}
- f^{\prime}\sin\frac{f}{2}\varSigma_{1}^{\prime} - 2\frac{a^{2}}{L^{2}}\cos\frac{f}{2}\varSigma_{1}
+ f^{\prime}\cos\frac{f}{2}\varSigma_{2}^{\prime} - 2\frac{\left( n_{a} - a \right)^{2}}{L^{2}}\sin\frac{f}{2}\varSigma_{2} \right], 
\label{eq:pemt_rr}\\
\delta T_{\theta\theta} &= - \frac{1}{g}a^{\prime}\left( a^{\prime}\mathcal{X} + \Dot{\varTheta}_{r} - \varTheta_{\theta}^{\prime} \right)
- L^{2}\Biggl\{
\mu\left[1 + 2\varPhi\left( 1 - \cos{f}\right) - 2\varSigma_{1}\cos\frac{f}{2} + 2\varSigma_{2}\sin\frac{f}{2} \right]
\notag\\
&- k\left[
\frac{f^{\prime 2}}{2}\left( \mathcal{X} - \varPhi \right) - \frac{v_{a}}{L^{2}}\varTheta_{\theta}
+ f^{\prime}\sin\frac{f}{2}\varSigma_{1}^{\prime} + 2\frac{a^{2}}{L^{2}}\cos\frac{f}{2}\varSigma_{1}
- f^{\prime}\cos\frac{f}{2}\varSigma_{2}^{\prime} + 2\frac{\left( n_{a} - a \right)^{2}}{L^{2}}\sin\frac{f}{2}\varSigma_{2}
\right]
\Biggr\}.
\label{eq:pemt_thetatheta}
\end{align}
have the only scalar mode fluctuations.
The two off-diagonal mixing components $(\mu,r)$ and $(\mu,\theta)$
\begin{align}
\delta T_{\mu r} &=
M\mathcal{E}_{\mu}\mathcal{L}_{\mathrm{brane}}
+ \frac{1}{g}\frac{a^{\prime}}{L^{2}}\left[ \partial_{\mu}\left( \Dot{\varTheta} - \varTheta_{\theta} \right) + \Dot{\varTheta}_{\mu} \right]
- k\partial_{\mu}\left( f^{\prime}\sin\frac{f}{2}\varSigma_{1} - f^{\prime}\cos\frac{f}{2}\varSigma_{2} \right), 
\label{eq:pemt_mur}\\
\delta T_{\mu\theta} &=
ML\mathcal{F}_{\mu}\mathcal{L}_{\mathrm{brane}}
+ \frac{1}{g}a^{\prime}\left[ \partial_{\mu}\left( \varTheta_{r} - \varTheta^{\prime} \right) - \varTheta_{\mu}^{\prime} \right]
- kv_{a}\left( \partial_{\mu}\varTheta + \varTheta_{\mu} \right)
\label{eq:pemt_mutheta}
\end{align}
have the vector and the scalar mode fluctuations.
Finally, the rest $(r,\theta)$ component
\begin{equation}
\delta T_{r\theta} =
L\varOmega\mathcal{L}_{\mathrm{brane}} + \frac{1}{g}\frac{a^{\prime 2}}{L}\varOmega
- k\left( f^{\prime}\sin\frac{f}{2}\Dot{\varSigma}_{1} - f^{\prime}\cos\frac{f}{2}\Dot{\varSigma}_{2} + v_{a}\varTheta_{r} \right)
\label{eq:pemt_rtheta}
\end{equation}
has the only scalar mode fluctuations.

\subsection{Matter field equations}
For the $\mathbb{C}P^{1}$ and $U(1)$ gauge fields,
one can obtain the perturbed equations of motion by inserting the perturbed fields (\ref{eqs:gi_pmatter}) into Eq.(\ref{eqs:per_eqs}).

For $a=1$ component of the evolution equation for the $\mathbb{C}P^{1}$ field,
the real part is
\begin{multline}
\frac{1}{M^{2}}\square\varSigma_{1} + \varSigma_{1}^{\prime\prime} + \frac{1}{L^{2}}\Ddot{\varSigma_{1}}
+ \left(4\Hat{m} + \Hat{\ell}\right)\left(\varSigma_{1}^{\prime} + \mathcal{X}f^{\prime}\sin\frac{f}{2}\right)
+ \left(\mathcal{X}f^{\prime\prime} + \frac{f^{\prime}}{2L}\Dot{\varOmega}\right)\sin\frac{f}{2}
\\
- \left(2\varPsi^{\prime} +\frac{\varPhi^{\prime} - \mathcal{X}^{\prime}}{2}\right)f^{\prime}\sin\frac{f}{2}
+ \left(\frac{f^{\prime 2}}{2}\mathcal{X} + \frac{2a^{2}}{L^{2}}\varPhi + \frac{2a}{L^{2}}\varTheta_{\theta}\right)\cos\frac{f}{2}
- \left(\frac{a^{2}}{L^{2}} - \frac{\mu}{k}\right)\varSigma_{1} = 0,
\label{apeq:p_cp1_1r}
\end{multline}
and the imaginary part is
\begin{multline}
\left(\frac{1}{M^{2}}\square\varTheta + \varTheta_{r}^{\prime} + \frac{1}{L^{2}}\Dot{\varTheta}_{\theta}\right)\cos\frac{f}{2}
+ \left(4\Hat{m} + \Hat{\ell}\right)\left(\varTheta_{r} + \frac{a}{L}\varOmega\right)\cos\frac{f}{2}
- \left(4\Dot{\varPsi} + \Dot{\mathcal{X}} - \Dot{\varPhi}\right)\frac{a}{L^{2}}\cos\frac{f}{2}
\\
+ \frac{1}{L}\left(a\varOmega^{\prime} + a^{\prime}\varOmega - a\varOmega\Hat{\ell}\right)\cos\frac{f}{2}
- \left(\varTheta_{r} + \frac{a}{L}\varOmega\right)f^{\prime}\sin\frac{f}{2}
- \frac{2a}{L^{2}}\Dot{\varSigma}_{1} = 0.
\label{eq:p_cp1_1i}
\end{multline}
For $a=2$ component,
the real part is
\begin{multline}
\frac{1}{M^{2}}\square\varSigma_{2} + \varSigma_{2}^{\prime\prime} + \frac{1}{L^{2}}\Ddot{\varSigma}_{2}
+ \left(4\Hat{m} + \Hat{\ell}\right)\left(\varSigma_{2}^{\prime} - \mathcal{X}f^{\prime}\cos\frac{f}{2}\right)
- \left(\mathcal{X}f^{\prime\prime} + \frac{f^{\prime}}{2L}\Dot{\varOmega}\right)\cos\frac{f}{2}
\\
+ \left(2\varPsi^{\prime} + \frac{\varPhi^{\prime} - \mathcal{X}^{\prime}}{2}\right)f^{\prime}\cos\frac{f}{2}
+ \left[\frac{f^{\prime 2}}{2}\mathcal{X} + \frac{2\left(n_{a} - a\right)^{2}}{L^{2}}\varPhi - \frac{2\left(n_{a} - a\right)}{L^{2}}\varTheta_{\theta}\right]\sin\frac{f}{2}
- \left[ \frac{\left(n_{a} - a\right)^{2}}{L^{2}} + \frac{\mu}{k} \right]\varSigma_{2} = 0,
\label{apeq:p_cp1_2r}
\end{multline}
and the imaginary part is
\begin{multline}
\left(\frac{1}{M^{2}}\square\varTheta + \varTheta_{r}^{\prime} + \frac{1}{L^{2}}\Dot{\varTheta}_{\theta}\right)\sin\frac{f}{2}
+ \left(4\Hat{m} + \Hat{\ell}\right)\left(\varTheta_{r} - \frac{n_{a} - a}{L}\varOmega\right)\sin\frac{f}{2}
- \left(4\Dot{\varPsi} + \Dot{\mathcal{X}} - \Dot{\varPhi}\right)\frac{n_{a} - a}{L^{2}}\sin\frac{f}{2}
\\
+ \frac{1}{L}\left[a^{\prime}\varOmega + \left(n_{a} - a\right)\left(\varOmega\Hat{\ell} - \varOmega^{\prime}\right)\right]\sin\frac{f}{2}
+ \left(\varTheta_{r} - \frac{n_{a} - a}{L}\varOmega\right)f^{\prime}\cos\frac{f}{2}
+ \frac{2\left(n_{a} - a\right)}{L^{2}}\Dot{\varSigma}_{2} = 0.
\label{eq:p_cp1_2i}
\end{multline}
There are the only scalar mode fluctuations in Eq.(\ref{apeq:p_cp1_1r})-(\ref{eq:p_cp1_2i}).

Similarly to the $\mathbb{C}P^{1}$ field,
one can obtain the evolution equations for the $U(1)$ gauge field.
The $(\mu)$ component of this equation has the vector mode sector
\begin{equation}
\frac{1}{M^{2}}\square\varTheta_{\mu}
+ \varTheta_{\mu}^{\prime\prime}
+ \frac{1}{L^{2}}\Ddot{\varTheta}_{\mu}
+ \left( 2\Hat{m} + \Hat{\ell} \right)\varTheta_{\mu}^{\prime}
- \frac{M}{L}a^{\prime}\left\{
\frac{1}{L}\Dot{\mathcal{E}}_{\mu} - \left[ \mathcal{F}_{\mu}^{\prime} + \left( \Hat{\ell} - \Hat{m} \right)\mathcal{F}_{\mu} \right]
\right\} = 8kg\varTheta_{\mu},
\label{eq:p_u1_mu_v}
\end{equation}
and the scalar mode sector
\begin{equation}
\varTheta^{\prime\prime}
- \varTheta_{r}^{\prime}
+ \left( 2\Hat{m} + \Hat{\ell} \right)\left( \varTheta^{\prime} - \varTheta_{r} \right)
+ \frac{1}{L^{2}}\left( \Ddot{\varTheta} - \Dot{\varTheta}_{\theta} \right)
= 8kg\varTheta.
\label{eq:p_u1_mu_s}
\end{equation}
In contrast to the component $(\mu)$,
the $(r)$ component
\begin{equation}
\frac{1}{M^{2}}\square\left( \varTheta_{r} - \varTheta^{\prime} \right)
+ \frac{1}{L^{2}}\left( \Ddot{\varTheta}_{r} - \Dot{\varTheta}_{\theta}^{\prime} \right)
+ \frac{a^{\prime}}{L^{2}}\left( 4\Dot{\varPsi} - \Dot{\mathcal{X}} - \Dot{\varPhi} \right)
- \frac{\varOmega}{L}v_{a} = 8kg\varTheta_{r}
\label{eq:p_u1_r}
\end{equation}
and the $(\theta)$ component
\begin{multline}
\frac{1}{M^{2}}\square\left( \varTheta_{\theta} - \Dot{\varTheta} \right)
+ \varTheta_{\theta}^{\prime\prime} - \Dot{\varTheta}_{r}^{\prime}
+ \left( 4\Hat{m} - \Hat{\ell} \right)\left( \varTheta_{\theta}^{\prime} - \Dot{\varTheta}_{r} \right)
- a^{\prime}\left( 4\varPsi^{\prime} - \mathcal{X}^{\prime} - \varPhi^{\prime} \right) + 2\mathcal{X}v_{a}
\\
= 8kg\left[ \varTheta_{\theta} - 2a\varSigma_{1}\cos\frac{f}{2} + 2\left( n_{a} - a \right)\varSigma_{2}\sin\frac{f}{2} \right]
\label{eq:p_u1_t}
\end{multline}
have the only scalar mode fluctuations.
\end{widetext}


\end{document}